\documentclass[preprints,article,accept,pdftex,moreauthors]{Definitions/mdpi}
\firstpage{1} 
\pubvolume{1}
\issuenum{1}
\articlenumber{0}
\pubyear{2022}
\copyrightyear{2022}
\datereceived{} 
\dateaccepted{} 
\datepublished{} 
\hreflink{https://doi.org/} 

\usepackage{amsmath}
\usepackage{caption}
\usepackage{subcaption}
\Title{Proton Electric Charge Radius from Lepton Scattering}

\TitleCitation{Proton Electric Charge Radius from Lepton Scattering}

\Author{
        Weizhi Xiong $^{1,\dagger}$
        , Chao Peng $^{2,\dagger}$
       }

\AuthorNames{Weizhi Xiong and Chao Peng}

\AuthorCitation{Xiong, W.; Peng, C.}

\address{%
$^{1}$ \quad Shandong University,  Qingdao, Shandong 266237, China; xiongw@sdu.edu.cn\\
$^{2}$ \quad Argonne National Laboratory, IL, USA; cpeng@anl.gov}

\firstnote{These authors contributed equally to this work.}

\abstract{Proton is a bound state of the strong interaction, governed by Quantum Chromodynamics (QCD). The electric charge radius of a proton, denoted by $r_{E}^{p}$, characterizes the spatial distribution of its electric charge carried by the quarks. It is an important input for bound-state Quantum Electrodynamic (QED) calculations of the hydrogen atomic energy levels. However, physicists have been puzzled by the large discrepancy between $r_{E}^{p}$ measurements from muonic hydrogen spectroscopy and those from $ep$ elastic scattering and ordinary hydrogen spectroscopy for over a decade. Tremendous efforts, both theoretical and experimental, have been dedicated to providing various insights into this puzzle, but certain issues still remain unresolved, particularly in the field of lepton scatterings. This review will focus on lepton-scattering measurements of $r_{E}^{p}$, recent theoretical and experimental developments in this field, as well as future experiments using this technique.}


\begin{document}

\section{Introduction}
Proton is the most stable hadron in the visible universe and is a  bound state of the strong interaction governed by Quantum Chromodynamics (QCD), with quarks and gluons as the fundamental degrees of freedom. The root-mean-square (rms) electric charge radius\footnote{For simplicity, the name ``proton charge radius'' is used for this quantity for the rest of the paper.} of the proton is an essential global quantity that characterizes the proton's electric charge size. This quantity is related to the spatial distribution of its charged constituents, \textit{i.e}., the quarks. However, precise theoretical calculations of the charge radius from the first principles are challenging as they require accurate knowledge of the proton's internal structure at the non-perturbative regime of QCD. In the past decade, the Lattice QCD method has shown promising developments and is expected to provide more precise \textit{ab-initio} calculations of the proton charge radius in the near future, which can be tested against experimental results. Additionally, the proton charge radius is an important input for bound-state Quantum Electrodynamics (QED) calculations of hydrogen atomic energy levels, and it is highly correlated with the Rydberg constant ($R_{\infty}$) -- one of the most precisely determined quantities in physics.

The proton charge radius can be determined from two well-established experimental methods~\cite{Gao:2021sml}. The first method involves measuring the proton electric form factor ($G_{E}^{p}$), which can be accessed through various types of experiments, including unpolarized electron-proton ($ep$) elastic scattering experiments~\cite{ A1:2010nsl, Xiong:2019umf, Mihovilovic:2019jiz}, elastic scattering experiments utilizing polarization degrees of freedom~\cite{Puckett:2010ac, Zhan:2011ji, Paolone:2010qc, Crawford:2006rz, Punjabi:2005wq}, and $e^{+}e^{-}$ annihilation experiments~\cite{Lin:2021xrc, BESIII:2021rqk}. The rms charge radius is determined from the slope of $G_{E}^{p}$ as the four-momentum transfer squared $Q^{2}$ approaches 0: 
\begin{equation}
  \langle {r_{E}^{p}}^{2} \rangle = - \frac{6}{G^p_E(0)}\frac{dG^p_E(Q^{2})}{dQ^{2}} {\biggr\rvert_{Q^{2} = 0} },
  \label{eq:radius_definition}
\end{equation}
where $G_{E}^{p} (0) = 1$ due to the charge normalization. Among these experimental results, the high-precision low-$Q^{2}$ $G_{E}^{p}$ data obtained from unpolarized $ep$ elastic scattering measurements play a critical role in the determination of $r_{E}^{p}$.

The second method for measuring $r_{E}^{p}$ is hydrogen spectroscopy~\cite{Beyer:2017gug, Fleurbaey:2018fih, Bezginov:2019mdi, Grinin2020, Brandt:2021yor}. This method exploits the shifts of the S-state energy levels caused by the proton's finite size. An S-state electron wave function is non-zero at the origin, thus, it can move inside the proton and experience a ``screening effect" due to the proton's charge. This effect slightly increases the electron energy level and is usually included in the Lamb shift. For ordinary hydrogen, the atomic energy level can be approximated as follows~\cite{Karr:2020wgh}:
\begin{equation}
  E_{n,l} \approx -\frac{R_{\infty}}{n^{2}} + \delta_{l,0}\frac{L_{1S}+a{r_{E}^{p}}^{2}}{n^{3}}.
  \label{eq:finite_size_effect}
\end{equation}
Here, $n$ and $l$ are the principle and angular momentum quantum numbers, respectively. $R_{\infty}$ is the Rydberg constant, $L_{1S}$ is the Lamb shift of 1S state of a point-like nucleus, and $a\approx1.56$ MHz$\cdot$fm$^{-2}$. Typically, two transition frequencies need to be measured to determine $R_{\infty}$ and $r_{E}^{p}$. However, for Lamb shift measurements such as the 2S$_{1/2}$ to 2P$_{1/2}$ transition~\cite{Bezginov:2019mdi}, the measurement of $R_{\infty}$ is not required, and thus this method provides an independent measurement of $r_{E}^{p}$.


Before 2010, the $r_{E}^{p}$ values obtained from modern $ep$ elastic scattering and hydrogen spectroscopy experiments were generally consistent with each other~\cite{Gao:2021sml}. According to CODATA-2010~\cite{Mohr:2012tt}, the $r_{E}^{p}$ values were $0.8758(77)$~fm and $0.895(18)$~fm from hydrogen spectroscopic and $ep$ elastic scattering experiments, respectively. Additionally, the A1 collaboration at Mainz Microtron (MAMI) extracted the proton charge radius using the unpolarized $ep$ elastic scattering in 2010, and reported $r_{E}^{p}=0.8791(79)$fm~\cite{A1:2010nsl}. Based on the above-mentioned results, CODATA-2010 determined the recommended value of $r_{E}^{p}$ as $0.8775(51)$fm. However, in 2010 and 2013, the CREMA collaboration published two results using a novel muonic hydrogen ($\mu$H) spectroscopy method~\cite{Pohl:2010zza, Antognini:2013txn} and found that the $r_{E}^{p}$ was significantly different from the CODATA-2010 recommended value. In this experiment, a muonic hydrogen atom was produced by replacing the atomic electron of an ordinary hydrogen atom with a muon, which is about 200 times heavier, resulting in a much smaller Bohr radius and greater sensitivity to the proton finite-size effect. The $\mu$H measurement obtained a value of $0.84184(67)$~fm, with an unprecedented 0.1\% precision. However, it was 4\% or 7$\sigma$ smaller than the CODATA-2010 recommended value, leading to the "proton charge radius puzzle". Since then, significant theoretical and experimental efforts have been devoted to understanding and resolving this discrepancy.

Theoretical uncertainties in the $\mu$H results are largely due to the contribution from Two-Photon Exchange (TPE) diagrams, which have been thoroughly investigated using various methods~\cite{Carlson:2011zd, Gorchtein:2013yga, Tomalak:2018uhr, Birse:2012eb, Peset:2015zga, Hill:2016bjv, Miller:2012ne}, including Lattice QCD~\cite{Fu:2022fgh}. Despite the consistency among these results, they cannot explain the significant discrepancy between the two $r_{E}^{p}$ values. To resolve this puzzle, new models have been proposed involving lepton-universality violation and new force carriers~\cite{Carlson:2015jba, Liu:2017htz, Liu:2018qgl, Bordes:2019qjk}. More specifically, an experimental approach involving dilepton photoproduction on a proton or deuteron target has been proposed and studied as a direct test of the lepton-universality at a facility that does not require muon beams \cite{Pauk:2015oaa, Carlson:2018ksu, Heller:2018ypa, Heller:2019dyv}. However, these new physics models lack sufficient experimental evidence to be supported at present.

The definition of $r_{E}^{p}$ has been comprehensively examined in a recent review by Peset, Pineda, and Tomalak~\cite{Peset:2021iul}, determining whether different experiments are measuring the same quantity within the context of effective field theory (EFT). While the same definition is achieved for muonic and ordinary hydrogen spectroscopic measurements, it is only achievable for lepton scattering in a rather restricted kinematic regime. Another problem is that ${G'}_{E}^{p}(0)$ becomes ill-defined once we consider electromagnetic corrections, which makes it inferred divergent, and scale and scheme dependent. Meanwhile, the extraction of $r_{E}^{p}$ from scattering experiments is still actively debated and investigated~\cite{Lee:2015jqa, Higinbotham:2015rja, Griffioen:2015hta, Bernauer:2016ziz, Yan:2018bez, Kraus:2014qua, Alarcon:2020kcz, Cui:2021vgm}, and no consensus has been reached on the best approach within the community. Recent progress in lattice QCD has significantly reduced the uncertainty of $r_{E}^{p}$~\cite{Hasan:2017wwt, Alexandrou:2018sjm, Jang:2019jkn, Shintani:2018ozy, Alexandrou:2020aja, Ishikawa:2021eut, Park:2021ypf, Djukanovic:2021cgp} and started to shed light on the puzzle. However, further improvements are needed to achieve a similar or better uncertainty compared to the results from empirical fits on experimental $G_{E}^{p}$ data.

Five ordinary hydrogen spectroscopic experiments have been published since the proton radius puzzle~\cite{Beyer:2017gug, Fleurbaey:2018fih, Bezginov:2019mdi, Grinin2020, Brandt:2021yor}. These new results have significantly reduced uncertainties through ground-breaking improvements, such as laser techniques and careful control of systematic uncertainties. While one of these experiments prefers the CODATA-2010 recommended value~\cite{Mohr:2012tt}, three of them, including the most precise result by Grinin \textit{et al}.~\cite{Grinin2020} and a Lamb shift measurement\cite{Bezginov:2019mdi}, favor the $\mu$H values. Interestingly, the latest ordinary hydrogen spectroscopic result by Brandt \textit{et al}.~\cite{Brandt:2021yor} measuring the transition frequency between 2S-8D states is about 3$\sigma$ away from both the muonic results and the CODATA-2010 value, indicating the need for further investigation from the spectroscopy community. For more information about hydrogen spectroscopy and recent progress, we would like to refer readers to recent reviews by Gao and Vanderhaeghen~\cite{Gao:2021sml}, Karr~\textit{et al.}~\cite{Karr:2020wgh}, Peset~\textit{et al.}~\cite{Peset:2021iul}, Antognini~\textit{et al.}~\cite{Antognini:2022xoo} and Pachucki~\textit{et al.}~\cite{Pachucki:2022tgl}.

\begin{figure}[H]
\centering
\includegraphics[width=13 cm]{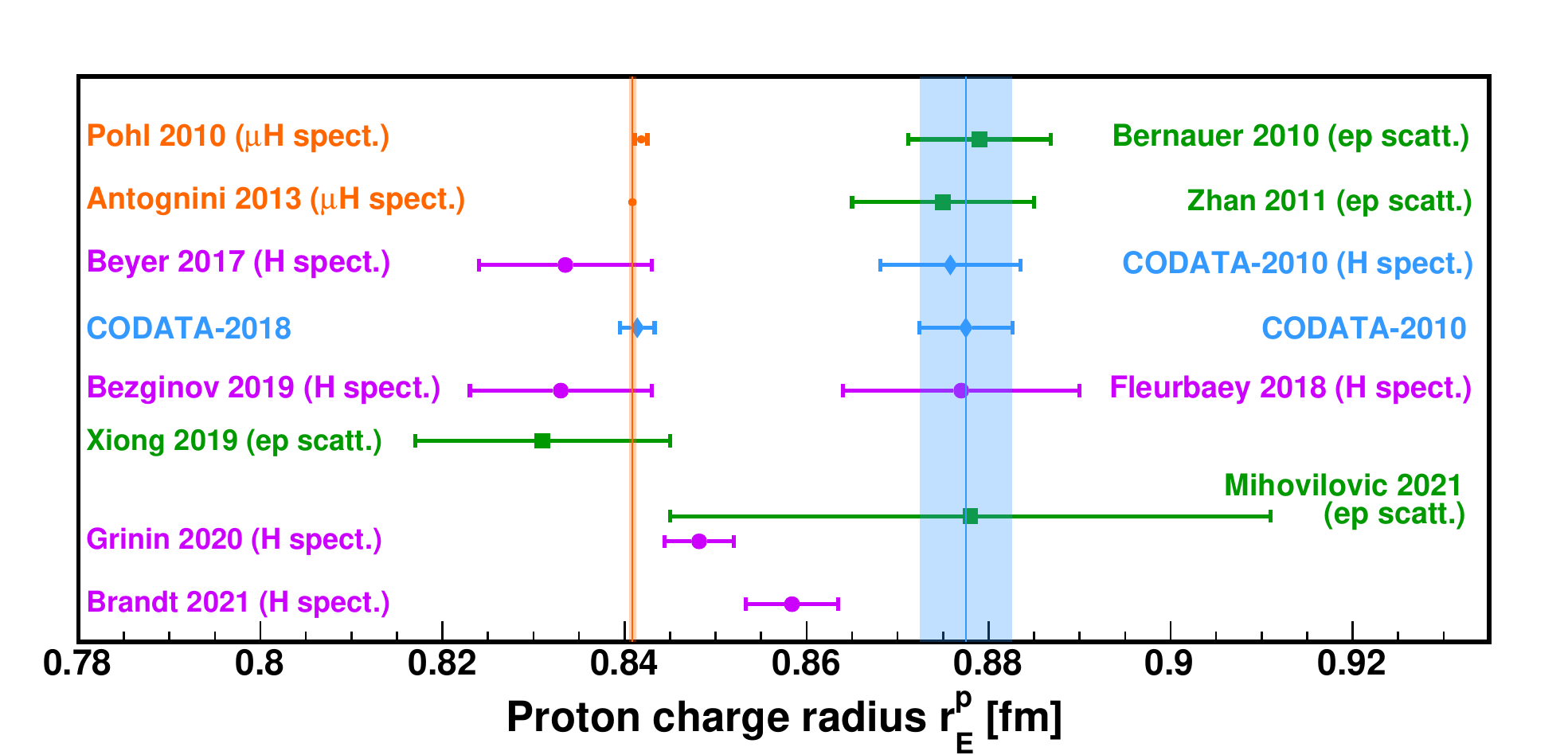}
\caption{The proton charge radius determined from $ep$ elastic scattering, hydrogen spectroscopic experiments, as well as world-data compilation from CODATA since 2010. The muonic spectroscopic measurements~\cite{Pohl:2010zza, Antognini:2013txn} are shown in orange dots, ordinary hydrogen spectroscopic results~\cite{Beyer:2017gug, Fleurbaey:2018fih, Bezginov:2019mdi, Grinin2020, Brandt:2021yor} are shown in purple dots, electron scattering measurements~\cite{A1:2010nsl, Zhan:2011ji, Mihovilovic:2019jiz, Xiong:2019umf} are shown in green squares, and blue diamonds show the CODATA compilations~\cite{Mohr:2012tt, Tiesinga:2021myr}.}
\label{fig:world_rp_data}
\end{figure}

On the lepton scattering side, the A1 collaboration at MAMI has developed novel experimental techniques to reach a lower $Q^{2}$ of 0.0013~GeV$^{2}$ by using the electron initial state radiation (ISR) technique\cite{Mihovilovic:2019jiz}. The collaboration also performed another experiment using a hydrogen gas jet target~\cite{A1:2022wzx}, which eliminates background from target cell windows. Although the results from these two experiments were limited by uncertainties, the development of these techniques will certainly benefit future experiments. The most impactful result from unpolarized lepton scattering since the puzzle is from the PRad experiment~\cite{Xiong:2019umf} at Thomas Jefferson National Accelerator Facility (TJNAF, a.k.a. Jefferson Lab or JLab). The experiment used a non-magnetic calorimetric setup with a windowless hydrogen-gas-flow target to simultaneously measure both elastic $ep$ and M{\o}ller ($ee$) scatterings, yielding a result of $r_{E}^{p} = 0.831(14)$~fm, which favors the $\mu$H values. However, a tension between the proton electric form factor data from PRad and those from the Mainz 2010 experiment, particularly in the $Q^{2}$ range between $0.01$~GeV$^{2}$ and $0.06$GeV$^{2}$, creates another puzzle that the scattering community needs to address in future experiments. Fig.~\ref{fig:world_rp_data} summarizes the recent $r_{E}^{p}$ measurements from both spectroscopy and electron scattering experiments.

After more than a decade, the proton charge radius puzzle remains an unresolved issue in the field. However, recent theoretical and experimental advancements offer promising prospects for deepening our understanding of this critical quantity. This review focuses on the progress made in the lepton scattering technique. We will begin with a brief introduction to the physics background, and then introduce recent scattering experiments performed since the discovery of the proton charge radius puzzle. We will then discuss recent advancements in calculating and extracting $r_{E}^{p}$ from form factor data, and then delve into the outstanding challenges in the field, followed by a brief overview of future elastic lepton-proton scattering experiments currently in preparation or data-taking phases.

\unskip

\section{Radius Extraction from Unpolarized Lepton-Proton Scattering Experiments}
\label{ch:FF_extraction}

\subsection{Empirical Fits of Electromagnetic Form Factors}
\label{sec:ff_fits}
The commonly used experimental method for measuring $G_{E}^{p}$ at low-$Q^{2}$ is the unpolarized lepton-proton scattering. Assuming that the lepton mass can be neglected, the elastic scattering cross section at Born level (single photon-exchange diagram) can be expressed by the Rosenbluth formula:
\begin{equation}
\frac{d\sigma}{d\Omega} = \left( \frac{d\sigma}{d\Omega} \right)_{\rm{Mott}} \frac{1}{1+\tau} \left[ ({G_{E}^{p}} (Q^{2}))^{2} + \frac{\tau}{\epsilon} ({{G_{M}^{p}} (Q^{2}))^{2}}\right],
\label{eq:diff_ep_cs}
\end{equation}
where $\tau = Q^{2}/(4M^{2})$ and $\epsilon = [1 + 2(1 + \tau)\tan^{2}(\theta/2)]^{-1}$, $M$ is the mass of a proton, and $\theta$ is the scattering angle of the lepton in the target rest frame. The Mott cross-section describes the scattering off a structure-less and spin-less proton:
\begin{equation}
\left( \frac{d\sigma}{d\Omega} \right)_{\rm{Mott}} = \frac{{\alpha}^{2} {\rm{cos}^{2}}\,\frac{\theta}{2}}{4E^{2} {\rm{sin}^{4}}\,\frac{\theta}{2}} \frac{E'}{E}~,
\label{eq:Mott_cs}
\end{equation}
where $\alpha$ is the fine-structure constant, and $E$ and $E'$ are the energies of the incoming and outgoing lepton, respectively. In the case where the lepton mass $m$ is not negligible, such as muon scattering, modifications to $\epsilon$ and Mott cross section ~\cite{Gramolin:2014pva, Tomalak:2014dja, Tomalak:2018jak} are needed:
\begin{equation}
\epsilon = \left[ 1 - 2(1+\tau) \frac{2m^{2} - Q^{2}}{4EE' - Q^{2}} \right]^{-1},
\end{equation}
\begin{equation}
{\left( \frac{d\sigma}{d\Omega} \right)_{\rm{Mott}}} = \frac{\alpha^{2}}{4E^{2}} \frac{1-Q^{2}/(4EE')}{Q^{4}/(4EE')^{2}} \frac{E |{\boldsymbol{\ell'}}|}{E' |{\boldsymbol{\ell}}|} \frac{M(E'^{2} - m^{2})}{MEE' + m^{2}(E' - E - M)},
\end{equation}
where $|{\boldsymbol{\ell}}|$ and $|{\boldsymbol{\ell}}'|$ are magnitudes of the three-momenta of the incident and scattered leptons, respectively.

To obtain the Born-level cross sections from a lepton scattering measurement, an unfolding process known as the ``radiative correction'' is required. This process typically consists of an ``internal'' correction and an ``external'' correction. The former relies on theoretical calculations of the contributions from higher-order Feynmann diagrams of the scattering process, and the latter corrects for the kinematic shifts of the detected particles due to their passage through materials. In addition, the overall radiative effects are expected to be smaller for muon scattering experiments, due to the muon's heavier mass compared to the electron. According to Ref.~\cite{Peset:2021iul}, future $\mu$p scattering experiments (MUSE and COMPASS) will require different modifications from the EFT in order to ensure a consistent definition of $r_{E}^{p}$.

In Eq.~\ref{eq:diff_ep_cs}, $G_{E}^{p}$ and $G_{M}^{p}$ respectively represent the proton electric and magnetic form factors, which contain information about the spatial distribution of the proton's charge and magnetization. These form factors are linear combinations of the Dirac ($F_{1}$) and Pauli ($F_{2}$) form factors:
\begin{equation}
  \begin{split}
  &G_{E}^{p}(Q^{2}) = F_{1}(Q^{2}) - \frac{Q^{2}}{4M^{2}}{\kappa}F_{2}({Q^{2}}),  \\
  &G_{M}^{p}(Q^{2}) = F_{1}(Q^{2}) + {\kappa}F_{2}({Q^{2}}),
  \end{split}
  \label{eq:dirac_pauli_ff}
\end{equation}
where $\kappa$ is the proton's anomalous magnetic moment.

It is noteworthy that the cross section given in Eq.~\ref{eq:diff_ep_cs} includes contributions from both $G_{E}^{p}$ and $G_{M}^{p}$ at the same $Q^{2}$. To extract each form factor from measured cross-sections, the Rosenbluth separation technique is commonly employed. This technique re-writes Eq.~\ref{eq:diff_ep_cs} into a reduced form:
\begin{equation}
\big( \frac{d\sigma}{d\Omega} \big)_{\rm{reduced}} =(1+\tau)\frac{\epsilon}{\tau}\frac{\left(\frac{d\sigma}{d\Omega}\right)_{ep}}{\left(\frac{d\sigma}{d\Omega}\right)_{\rm{Mott}}} = (G_{M}^{p} (Q^{2}))^{2} + \frac{\epsilon}{\tau} (G_{E}^{p} (Q^{2}))^{2}.
\label{eq:rosenbluth_reduced_cs}
\end{equation}
By measuring cross sections at different values of $\epsilon$ but the same $Q^{2}$, $G_{E}^{p}$ and $G_{M}^{p}$ can be determined separately by the slope and intersection of a linear fit to the reduced cross-section data. Alternatively, one can parameterize $G_{E}^{p}(Q^{2})$ and $G_{M}^{p}(Q^{2})$ using different functions or models and directly fit the parameters to the measured cross sections. The analysis of the Mainz 2010 data~\cite{A1:2010nsl} utilized this approach. In addition, if only the electric form factor is of interest, the contribution from the magnetic form factor can be estimated and subtracted using models or parameterizations, allowing a direct extraction of $G_{E}^{p}$ from the cross-section data. However, this approach is only applicable in the low-$Q^{2}$ and extreme-forward angular region, where the kinematic factor $\tau/\epsilon$ significantly suppresses the $G_{M}^{p}$ contribution. The PRad experiment~\cite{Xiong:2019umf}, which covered $2.1\times10^{-4} < Q^{2} < 5.8\times10^{-2}$ GeV$^2$ and $0.7^{\circ} < \theta < 7.0^{\circ}$, utilized this approach.

Once the form factors are extracted from the cross sections, the slope of the form factor data at $Q^{2} = 0$ can be used to obtain the radius, as shown in Eq.~\ref{eq:radius_definition}. However, it is not experimentally feasible to directly measure $Q^{2} \to 0$, so experiments need to use physics-based models or empirical fits to describe the form factor data measured at finite $Q^{2}$, and then extrapolate to $Q^{2} = 0$ for radius extraction. As a result, a systematic uncertainty associated with different choices of fitting functions or models is often inevitable. Currently, no consensus in the community has been reached on the best model or empirical fit, and the choice is highly dependent on the kinematic range of the experiment. However, including high-precision form factor data with lower $Q^{2}$ can reduce this systematic uncertainty. In addition, pseudo-data methods~\cite{Yan:2018bez,  Kraus:2014qua} can test the robustness of different parameterizations and models, providing a valuable and effective means of handling this uncertainty. Some popular empirical fits include, but are not limited to, the multi-parameter rational function of $Q^{2}$ (rational ($N, M$)):
\begin{equation}
G^p_E(Q^{2}) = \frac{ 1+ \sum_{i=1}^{N} p_{i}^{a}Q^{2i}}{ 1 + \sum_{j=1}^{M} p_{j}^{b}Q^{2j}},
\label{eq:rational_fitter}
\end{equation}
the multi-parameter polynomial expansion in $Q^{2}$:
\begin{equation}
G^p_E(Q^{2}) = 1 + \sum_{i=1}^{N} p_{i}Q^{2i},
\label{eq:poly_fitter}
\end{equation}
and the multi-parameter polynomial expansion in $z$:
\begin{equation}
  \begin{split}
  &G^p_E(Q^{2}) = 1 + \sum_{i=1}^{N}  p_{i}z^{i},  \\
  &z = \frac{\sqrt{T_{c} + Q^{2}} - \sqrt{T_{c} - T_{0}}}{\sqrt{T_{c} + Q^{2}} + \sqrt{T_{c} - T_{0}}},
  \end{split}
  \label{eq:z_expansion_fitter}
\end{equation}
where $T_{c} = 4m_{\pi}^{2}$, $m_{\pi}$ is the pion mass, and $T_{0}$ is a free parameter representing the point that is mapping onto $z = 0$. In addition, a simple dipole fitter is commonly used to flatten the $G_E^p$ data over a large $Q^2$ range, as
\begin{equation}
G^p_E(Q^{2}) = \big( 1 + \frac{Q^{2}}{p_{1}} \big)^{-2}.
\label{eq:dipole_fitter}
\end{equation}
With $p_{1} = 0.71$~GeV$^{2}$, the dipole fitter is referred to as the standard dipole form factor. From Eq.~\ref{eq:rational_fitter} to Eq.~\ref{eq:dipole_fitter}, the fitting parameters $p_{i}$ are obtained from a fit to the experimental data. It is worth noting that fit to the experimental data often suffers from normalization uncertainties due to multiple experimental settings and limited knowledge of the absolute luminosity for each setting. A floating normalization parameter can be introduced to account for these uncertainties, given by
\begin{equation}
	f(Q^{2}) = n\mathcal{G}^p_E(Q^{2}),
\end{equation}
where $f$ is the final functional form, $n$ is the floating normalization parameter, and $\mathcal{G}^p_E(Q^{2})$ is the original functional form.

\subsection{Moments of Transverse Charge Density}
\label{sec:ff_fits}
The proton charge radius was defined as the second moment of the three-dimensional charge distribution of a proton in the Breit frame (BF)~\cite{Ernst:1960zza, Sachs:1962zzc}. However, this definition of the proton charge distribution does not contain proper relativistic contents~\cite{Miller:2018ybm, Jaffe:2020ebz}. One way to incorporate relativistic corrections is to define a two-dimensional transverse density in the infinite-momentum frame (IMF) using the light-front formalism~\cite{Miller:2018ybm}
\begin{equation}
{\rho}(b) = \frac{1}{2\pi}\int_{0}^{\infty}QF_{1}(Q^{2}){J}_{0}(Qb)dQ,
\label{eq:true_density}
\end{equation}
\begin{equation}
F_{1}(Q^{2}) = 2{\pi}\int_{0}^{\infty}b{\rho}(b)J_{0}(Qb)db,
\label{eq:inverse_true_density}
\end{equation}
where $b$ is the impact parameter, and $J_{0}$ is the cylindrical Bessel function. If we expand $J_{0}(Qb)$, Eq.~\ref{eq:inverse_true_density} can be expressed as
\begin{equation}
F_{1}(Q^{2}) \approx 1 - \frac{Q^{2}}{4} {\langle b^{2} \rangle} + \frac{Q^{4}}{64} {\langle b^{4} \rangle} - \dots,
\label{eq:inverse_true_density_expansion}
\end{equation}
where $\langle b^{2n} \rangle$ are the moments of the transverse charge density and are Lorentz invariant. The second moment or the root-mean-square transverse radius is then
\begin{equation}
{\langle b^{2} \rangle} = 2{\pi} \int_{0}^{\infty} b^{3}{\rho}(b)db.
\label{eq:second_moment_transverse_charge_density}
\end{equation}
If we take a Taylor expansion of $F_{1}(Q^{2})$ at $Q^{2} = 0$ and compare it to Eq.~\ref{eq:inverse_true_density_expansion}, another definition of transverse radius is obtained:
\begin{equation}
\langle b^{2} \rangle = - \frac{4}{F_{1}(0)}\frac{dF_1(Q^{2})}{dQ^{2}} {\biggr\rvert_{Q^{2} = 0} }.
\label{eq:transverse_radius_slope}
\end{equation}
One immediately notes that the transverse radius is also related to the slope of $F_{1}$ at $Q^{2} = 0$, similar to the definition of the proton charge radius in Eq~\ref{eq:radius_definition}. In fact, the connection between $G_E$ in BF and $F_1$ in IMF was shown in the work of Rinehimer and Miller \cite{Rinehimer:2009yv}. Combining Eq~\ref{eq:radius_definition}, Eq.~\ref{eq:transverse_radius_slope}, and the derivative of Eq.~\ref{eq:dirac_pauli_ff} at $Q^{2} = 0$ leads to a simple model-independent relation between $\langle b^{2} \rangle$ and $\langle {r_{E}^{p}}^{2} \rangle$:
\begin{equation}
\langle {r_{E}^{p}}^{2} \rangle = \frac{3}{2} \left( \langle b^{2} \rangle + \frac{\kappa}{M^{2}}\right).
\label{eq:rp_b_relation}
\end{equation}
This offers an alternative method to extract the proton charge radius $r_{E}^{p}$, which has already been utilized in some recent analyses~\cite{Atac:2020hdq, Gramolin:2021gln}.

Besides the light-front formalism, various approaches have been explored to properly define the relativistic charge density. In particular, it is important to establish a clear relationship between the definition and the rest-frame charge distribution which is directly probed in fixed-target scattering experiments. Freeze and Miller recently demonstrated that using the light-front formalism with tilted light-front coordinates leads to a frame-independent definition of the charge density, which characterizes the internal charge distribution of the proton~\cite{Freese:2023jcp}. Lorc{\'e} and Chen developed a quasi-probabilistic interpretation using the quantum phase-space formalism and provided a natural interpolation between the BF and IMF charge distributions, avoiding ambiguities associated with the BF charge distribution and relativistic corrections~\cite{Lorce:2020onh, Chen:2022smg, Chen:2023dxp}. In particular, the physical interpretation of the nucleon polarization and magnetization spatial distributions was recently formulated systematically in Ref.~\cite{Chen:2023dxp}. Epelbaum \textit{et al.} introduced a definition of local charge densities based on sharply localized, spherically symmetric wave packets and showed that the radial moments of the charge distribution are independent of the choice of frames~\cite{Epelbaum:2022fjc}. Li \textit{et al.} revisited the macroscopic field theory and found that the Sachs charge distribution and the light-front charge distribution are different types of multipole moment expansions, which can be measured at any frame but require convergence on the expansion~\cite{Li:2013pys}.

\section{Recent Progress from Electron Scattering Experiments}
In this section, we will provide a brief overview of three unpolarized $ep$ elastic scattering experiments conducted after 2010. These experiments are the Initial-State Radiation experiment~\cite{Mihovilovic:2019jiz} and the jet-target experiment~\cite{A1:2022wzx}, both performed at MAMI, and the PRad experiment~\cite{Xiong:2019umf}, conducted at Jefferson Lab.

\subsection{Initial-State Radiation Experiment at Mainz}

\begin{figure}
    \centering
	\includegraphics[width=10cm]{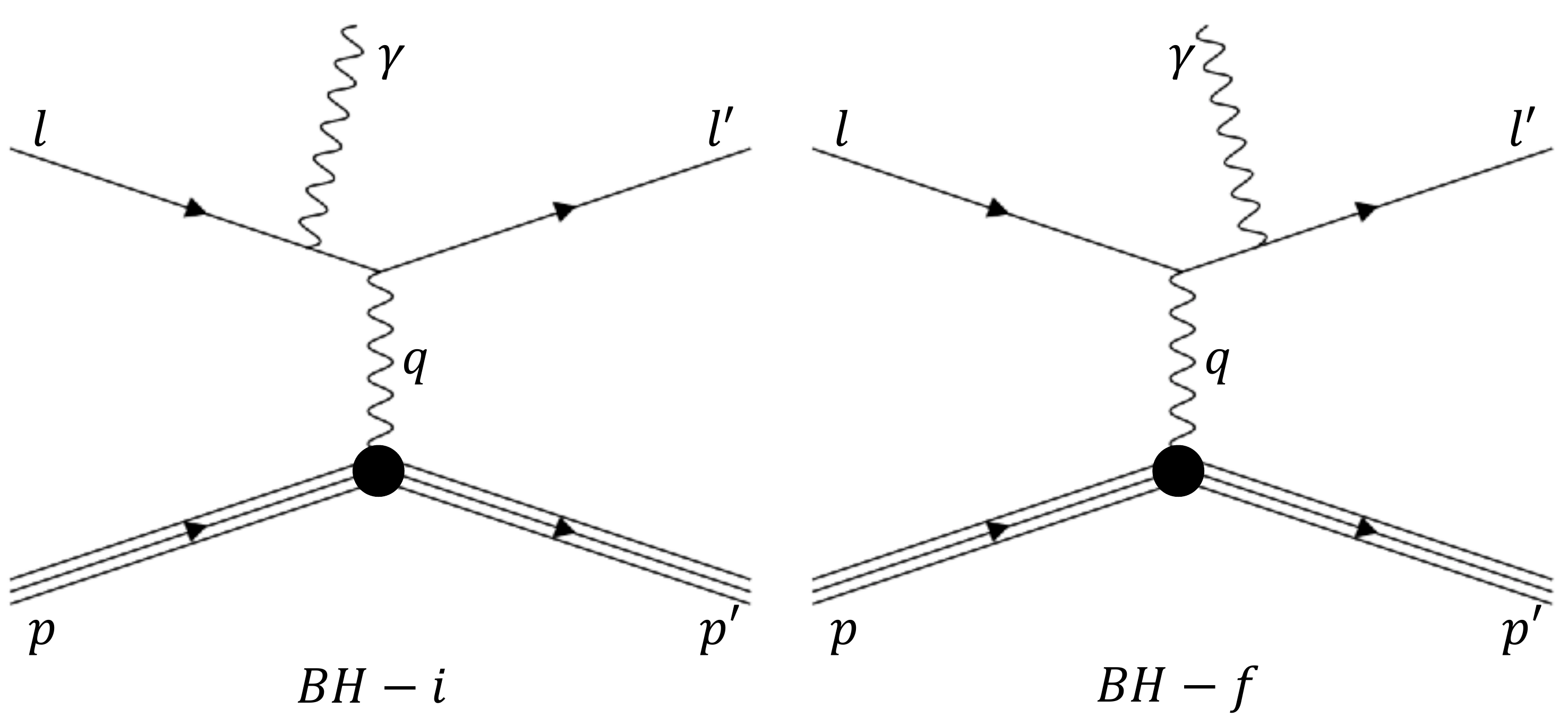}
	\caption{Feynman diagrams for the initial state (left panel, BH-i), and the final state (right panel, BH-f) Bethe-Heitler processes.}
	\label{fig:BH_diagram}
\end{figure}

The initial-state radiation (ISR) experiment at Mainz collected elastic $ep$ scattering data using the initial radiation technique~\cite{Mihovilovic:2019jiz}. The lowest $Q^2$ accessible to an elastic $ep$ scattering experiment is determined by the lowest beam energy and the most forward electron-scattering angle, which are often limited by the accelerator facility and experimental apparatus. This experiment has further lowered the reachable $Q^2$ of the existing elastic $ep$ cross-section data of Mainz down to 0.001 GeV$^2$ by determining the contribution from the initial-state Bethe-Heitler process (labeled as BH-i in Fig.~\ref{fig:BH_diagram}) within the radiative tail of the elastic peak.

The Mainz ISR experiment in 2013 employed the same three spectrometer setup as the Mainz 2010 experiment to measure the radiative tail. They used three different beam energies, namely $E_{beam}$ = 195, 330, and 495 MeV, with a beam current ranging from 10 nA to 1 $\mu$A, along with a five-centimeter-long liquid hydrogen target. The full radiative tail was scanned at a scattering angle of 15.21$^{\circ}$ by adjusting spectrometer B, while spectrometer A was fixed at a specific setting to monitor the luminosity.

The measured radiative tail data were analyzed using a dedicated Monte-Carlo simulation to separate the contributions from the BH-i and BH-f diagrams (see Fig.~\ref{fig:BH_diagram}). External radiative effects were corrected using Mo and Tsai formalism~\cite{Mo:1968cg}, while the collisional loss was fit to a Landau distribution. The thickness of the materials was determined by measuring the elastic electron scattering off residual nitrogen/oxygen gas using spectrometer A. The ISR technique yielded a total of 25 data points within $0.001 \le Q^2 \le 0.017$ GeV$^2$/c$^2$. A polynomial fit, with 3 free parameters and the normalization factor for each data set, was used to extract the proton charge radius, which was found to be
\begin{equation}
r_{E}^{p} = 0.873 \pm 0.017_{\rm{stat.}} \pm 0.059_{\rm{syst.}} \pm 0.003_{\rm{mod.}}\rm{~fm}.
\label{eq:ISR_result1}
\end{equation}
Here, the ``mod'' term represents the uncertainty associated with higher moments in parameterizing the proton electric form factor. An alternative approach was also used, in which the data were compared with simulations based on a polynomial parameterization of the form factors, with the proton charge radius as the only free parameter. This approach discarded the data with $E_{beam}$ = 195 MeV due to their limited $Q^2$ coverage and used the data sets with $E_{beam}$ = 330 and 495 MeV. The final result was given as a weighted average between the fits for the two energy settings:
\begin{equation}
r_{E}^{p} = 0.878 \pm 0.011_{\rm{stat.}}\pm 0.031_{\rm{syst.}}\pm 0.002_{\rm{mod.}}\rm{~fm}.
\label{eq:ISR_result2}
\end{equation}
However, the radii extracted from the two individual settings exhibit a 2.5$\sigma$ tension between each other, with the higher energy setting favoring $r_p \approx 0.84$ fm and the other favoring $r_p \approx 1.0$ fm~\cite{Mihovilovic:2019jiz}.

\subsection{Proton Charge Radius Experiment at JLab}
\begin{figure}
    \centering
	\includegraphics[width=0.95\textwidth]{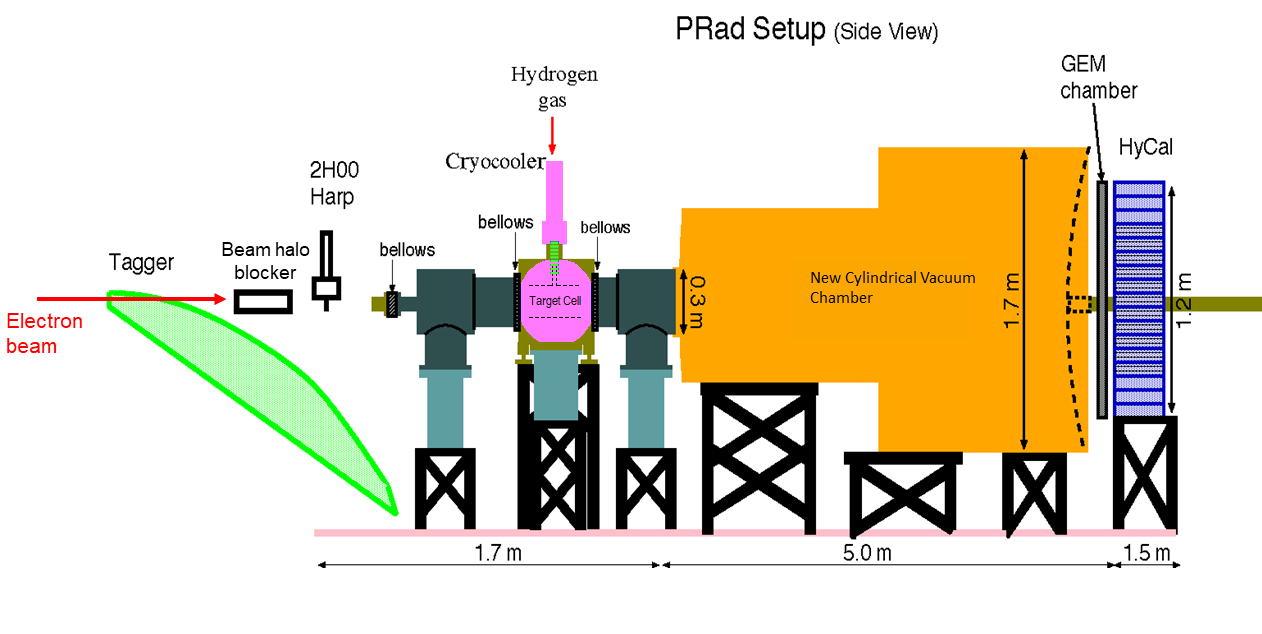}
	\caption{The schematic of the PRad experiment.}
	\label{fig:prad_layout}
\end{figure}

The Proton Charge Radius Experiment~\cite{Xiong:2019umf} at Jefferson Lab was performed in Hall B by the PRad collaboration in 2016. It measured the differential cross sections of $ep$ elastic scattering in a scattering angle range of 0.7$^{\circ}$ to 7.0$^{\circ}$ at beam energies of 1.101 GeV and 2.143 GeV. As depicted in Fig.~\ref{fig:prad_layout}, the PRad experiment utilized a calorimetric technique to measure the unpolarized $ep$ elastic scattering, which, in conjunction with a novel windowless hydrogen-gas-flow target \cite{Pierce:2021vkh}, yielded a set of systematic uncertainties distinct from those of other modern scattering experiments.

The PRad detector system was comprised of a Hybrid Calorimeter (HyCal) and a plane of Gas Electron Multipliers (GEM). It achieved relative energy resolutions of $2.4\%/\sqrt{E}$ (central PbWO$_4$ region) and $6.2\%/\sqrt{E}$ (outer Pb-Glass region), and a position resolution of about 79 $\mu$m (single GEM plane)~\cite{Bai:2020wao}. The system had a broad geometrical acceptance that covered two orders of magnitude in $Q^2$ with a single beam energy setting, which substantially decreased the systematic uncertainties associated with the normalization factors for combining data from different experimental configurations.

The PRad experiment utilized a unique windowless hydrogen-gas-flow target~\cite{Pierce:2021vkh} contained in a high-vacuum chamber that was connected to the beamline. The target cell was consistently injected with a flow of cryogenic H$_2$ gas at $T_0 \approx 20$ K, which was then pumped out through the vacuum pumps attached to the target chamber. The hydrogen gas was uniformly distributed along the four-centimeter-long cell, with a small fraction of residual gas extending through the aperture at both ends of the target cell. The target achieved an areal thickness greater than $2 \times 10^{18}$ atoms/cm$^2$. This windowless target largely eliminated the typical background source from scattering measurements, \textit{i.e.}, the target window, at the cost of additional background from the residual gas and difficulties in precisely determining the target thickness. The systematic uncertainties associated with the former were minimized by subtracting the data from empty target runs. Those associated with the latter were suppressed by normalizing the elastic $ep$ yield to that from the well-known M{\o}ller process.

During the PRad experiment, scattered electrons from both $ep$ elastic and M{\o}ller scatterings were simultaneously measured. These two types of events were easily separated by their energies within the kinematic coverage of the experiment. The experimental yields of $ep$ elastic scattering were normalized to those of the M{\o}ller process, cancelling out the luminosity at first order in this ratio measurement. To extract the proton's electric form factors, an iterative process of radiative corrections based on a full GEANT4 \cite{GEANT4:2002zbu} simulation and complete first-order calculations beyond the ultra-relativistic approximation (URA) \cite{Akushevich:2015toa} was applied to the data. The extracted $Q^2$ ranged from $2.1 \times 10^{-4}$~GeV$^2$ to $5.8 \times 10^{-2}$~GeV$^2$. The measured cross sections include the contributions from both electric and magnetic form factors, in which $G_{E}^{p}$ is the predominant term at low $Q^2$, according to Eq.~\ref{eq:rosenbluth_reduced_cs}. In the PRad case, the $G_{M}^{p}$ contribution was negligible at the lowest $Q^2$ data achievable with the 1.1 GeV beam, ranging from 0.015\% to 0.06\%. At higher $Q^2$, the $G_{M}^{p}$ contribution was estimated using the Kelly parameterization~\cite{Kelly:2004hm}, and its associated systematic uncertainty for the extracted $G_{E}^{p}$ was less than 0.3\%.

The PRad collaboration utilized a Rational (1, 1) fit (see Eq.~\ref{eq:rational_fitter}) to extract the proton charge radius from the measured electric form factors, resulting in a value of \cite{Xiong:2019umf}
\begin{equation}
r_p = 0.831 \pm 0.007_{\rm{stat}} \pm 0.012_{\rm{syst}} \rm{fm}.
\end{equation}
This value is consistent with the $\mu H$ spectroscopy results from the CREMA collaboration\cite{Pohl:2010zza, Antognini:2013txn}. Despite the fact that the systematic uncertainties of the PRad experiment are very different than those of the other modern scattering experiments, PRad value is in direct conflict with the proton charge radius extracted from the Mainz 2010 experiment \cite{A1:2010nsl} or the JLab recoil polarization experiment \cite{Zhan:2011ji}. Moreover, a discrepancy in the $G_{E}^{p}$ values between PRad and Mainz 2010 was observed, particularly in the range of $0.01 < Q^2 < 0.06$ GeV$^2$, which will be discussed in detail in Section~\ref{sec:remaining_issue}.

\subsection{Jet Target Experiment at Mainz}
\label{ch:mainz_jet_exp}
The Mainz jet-target experiment re-measured the proton electric form factor at low-$Q^2$ from 0.01 to 0.045 GeV$^2$~\cite{A1:2022wzx} and investigated the data tension between PRad and Mainz 2010. This experiment utilized the A1 multi-spectrometer facility at MAMI and measured elastic $ep$ scattering with a novel cryogenic supersonic gas jet target~\cite{A1:2021njh}, with molecular hydrogen gas cooled to cryogenic temperatures. It was conducted with $E_{beam} = 315$ MeV and a beam current of $20$~$\mu$A. Unlike the gas-flow target used by PRad, this target was designed to be compact (about 1 mm in length) and achieved a comparable areal thickness of $10^{18}$ atoms/cm$^2$ at a nominal gas-flow rate of $q_V = 2400$ l$_n$/h and temperature of $T_0 = 40$ K. The target system was placed in a high-vacuum scattering chamber, where the cryogenic gas was compressed through a vertical convergent-divergent nozzle. The gas jets interacted with the horizontally traversing electron beam at a certain angle and were eventually disposed into an aligned catcher placed a few millimeters away. To reject the background from the beam halo hitting the nozzle-catcher structure, a veto system was used, which included a pair of tungsten collimators and a double-arm scintillating detector mounted upstream of the target.

The experiment collected elastic $ep$ scattering data covering a range of scattering angles from 15$^{\circ}$ to 40$^{\circ}$ and reaching a low $Q^2$ of 0.01 GeV$^{2}$. While the collimators effectively removed beam-halo-associated backgrounds, the veto detectors failed to function under the operating conditions, leading to some residual backgrounds from the halo. To further investigate the background, a low flow-rate run at $q_V = 50 $l$_n$/h was conducted since a zero flow-rate run was technically unfeasible. The collected data were subjected to radiative correction using a Monte-Carlo simulation with a re-developed generator, adapted from the OLYMPUS experiment \cite{Schmidt:2016fsl}. The experiment could not determine the absolute luminosity because it was impossible to directly measure the density distribution of the target that overlaps with the beam. Therefore, the $ep$ scattering at 30$^{\circ}$ was employed as a luminosity monitor in the experiment, and the global normalization of the luminosity was set as a free parameter for fit.

\begin{figure}
    \centering
	\includegraphics[width=0.85\textwidth]{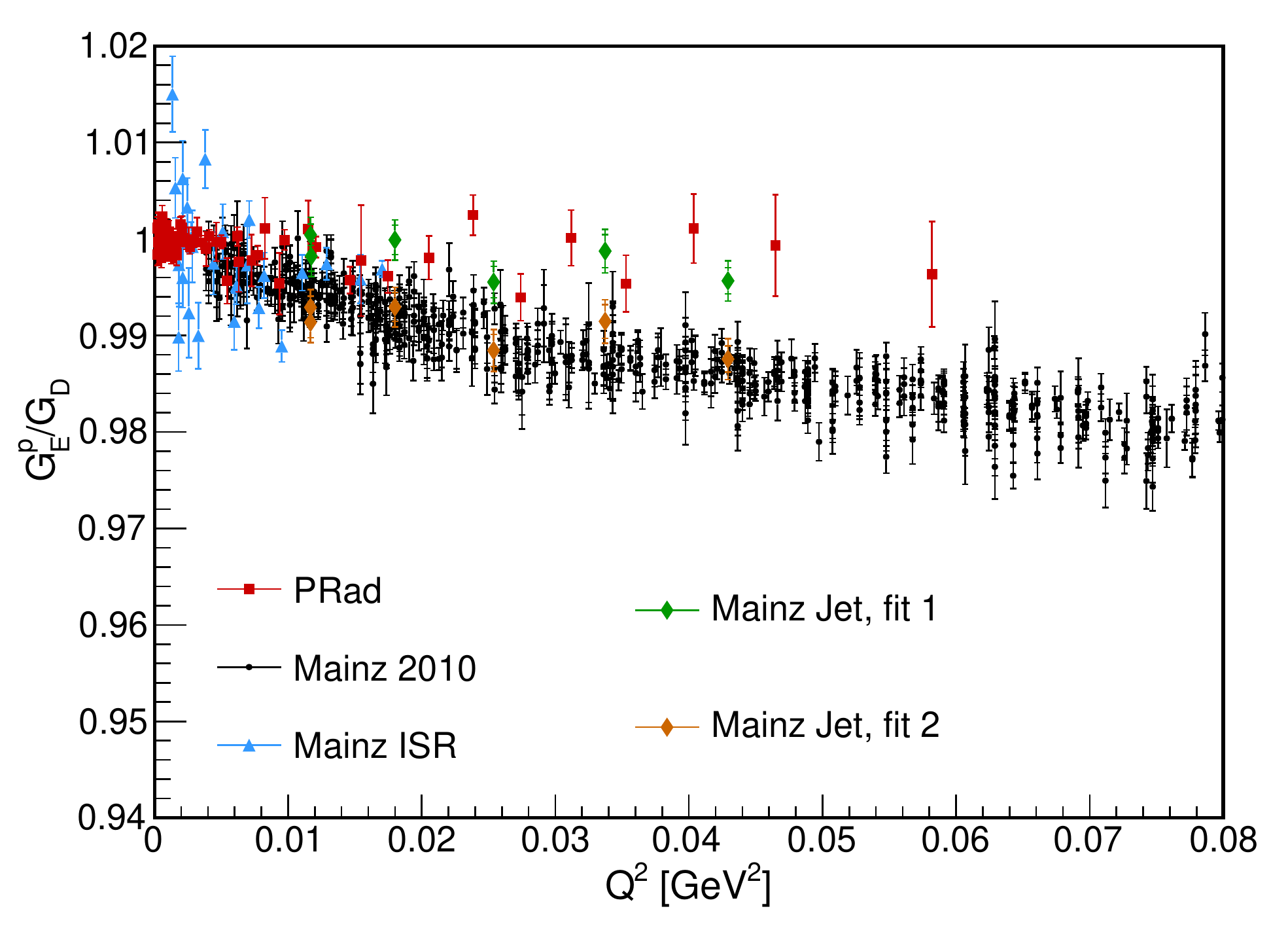}
	\caption{The proton electric form factor measured in the Mainz jet target experiment~\cite{A1:2022wzx}. The global luminosity was fit to PRad (Mainz jet, fit 1) and Mainz A1 (Mainz jet, fit 2) separately. Also shown are the $G_{E}^{p}$ data, normalized by the standard dipole form factor, from the Mainz ISR~\cite{Mihovilovic:2019jiz}, PRad~\cite{Xiong:2019umf}, and the Mainz 2010~\cite{A1:2010nsl} experiments.}
	\label{fig:recent_ff}
\end{figure}

In Fig.~\ref{fig:recent_ff}, the measured $G_{E}^{p}$ in this experiment is presented, with the $G_{M}^{p}$ contribution estimated by Kelly's parameterization \cite{Kelly:2004hm}. The global luminosity was fitted using both PRad's rational (1,1) and Mainz's polynomial parameterizations for $G_{E}^{p}$, separately. The experimental data obtained using the jet target are consistent with both PRad and Mainz parameterizations, with a slightly better $\chi^2$ from PRad's rational (1,1) fit. However, due to the limited statistical uncertainty, this experiment is unable to resolve the data tension between PRad and Mainz for $0.01 < Q^2 < 0.06$ GeV$^{2}$. Nevertheless, this experiment has demonstrated the feasibility of deploying a windowless jet target with high-resolution spectrometers, and it provides a set of systematic uncertainties that differ from those obtained with a traditional target in previous spectrometer experiments. This technique is expected to be beneficial for certain future experiments, such as the MAGIX experiment at MESA~\cite{`A1:2021njh}.

\section{Recent Re-Analyses and Lattice QCD Calculations}
The recent experimental progress has been accompanied by advancements in re-analyses of world form factor data and Lattice QCD calculations, which remain an active research field. These theoretical efforts offer unique insights into the proton charge radius puzzle. In this section, we will give a brief overview of the recent developments in these two areas. Some of the earlier studies are discussed in the review by Gao and Vanderhaeghen~\cite{Gao:2021sml}, Peset~\textit{et al.}~\cite{Peset:2021iul}, as well as the recent conference paper by Mei{\ss}ner~\cite{Meissner:2022rsm}.

\subsection{Re-analysis of Form Factor Data}

\begin{figure}[H]
\centering
\includegraphics[width=13 cm]{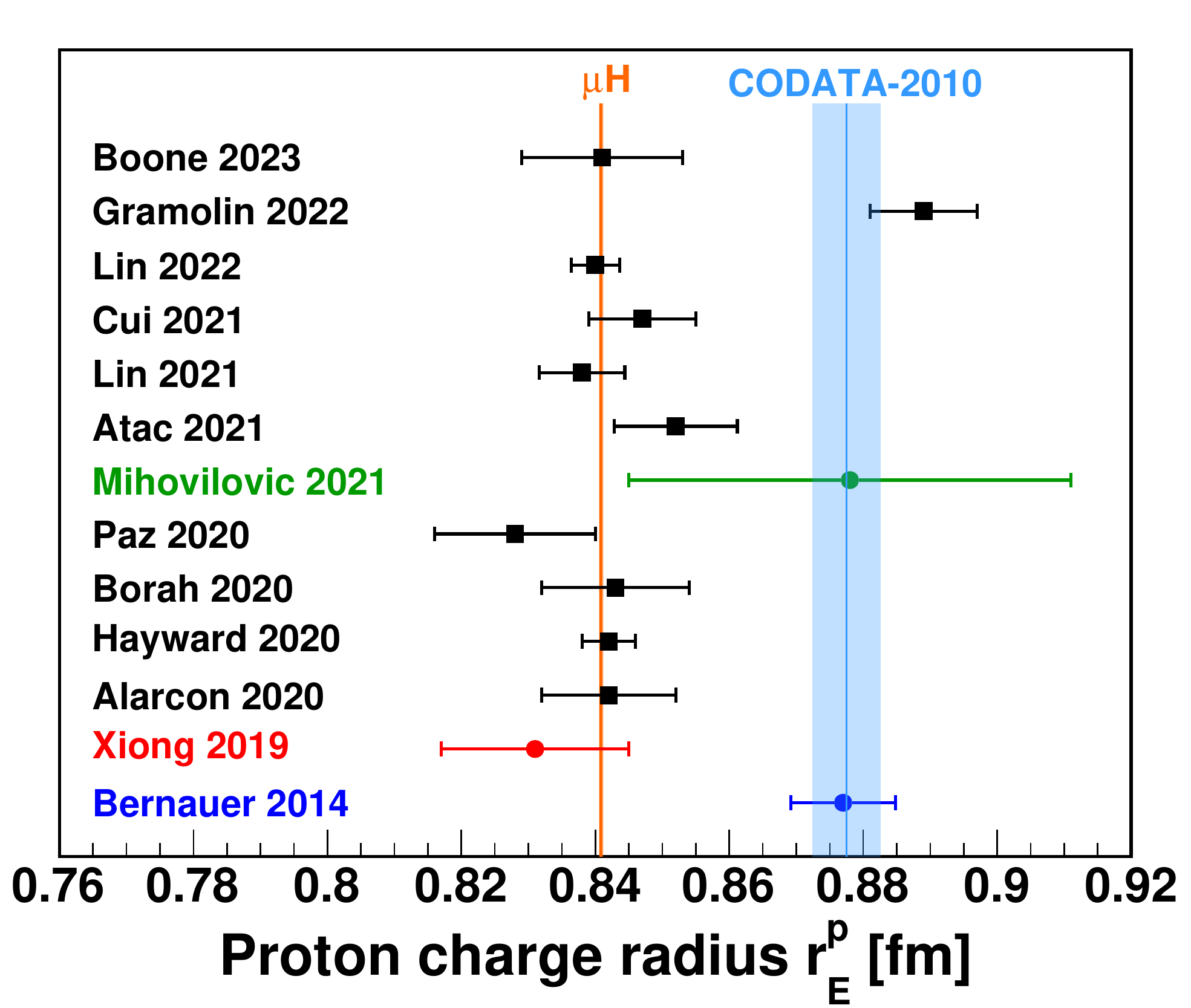}
\caption{The proton charge radius extracted from a number of recent re-analyses~\cite{Alarcon:2020kcz, Hayward:2018qij, Borah:2020gte, Paz:2020prs, Atac:2020hdq, Lin:2021umz, Cui:2021vgm, Lin:2021xrc, Gramolin:2021gln, Boone:2023ryu} (black squares), as well as those from the original analyses of three recent unpolarized $ep$ elastic scattering experiments (colored dots). The dark blue dot is for the Mainz 2010 experiment~\cite{A1:2010nsl, A1:2013fsc}, the red dot is for the PRad experiment~\cite{Xiong:2019umf}, and the dark green dot is for the Mainz ISR experiment~\cite{Mihovilovic:2019jiz}. The orange line and band (barely visible due to small uncertainty) correspond to the 2013 $\mu$H spectroscopic measurement~\cite{Antognini:2013txn}, and the light blue line and band correspond to the CODATA-2010 recommended value~\cite{Mohr:2012tt}. }
\label{fig:rp_reanalysis}
\end{figure}

In this subsection, our attention is directed towards the recent re-analyses of global data following the PRad experiment. These investigations can be classified into two categories: radius extraction utilizing physics-driven models, and those utilizing empirical fits or statistical methods. Fig.~\ref{fig:rp_reanalysis} presents a compilation of proton charge radii obtained from several recent studies.

Alarc\'{o}n, Higinbotham, and Weiss~\cite{Alarcon:2020kcz} developed a novel theoretical framework that combines dispersion analysis and Chiral Effective Field Theory (DI$\chi$EFT)~\cite{Alarcon:2018zbz, Alarcon:2018irp, Alarcon:2017lhg}. In this model, the electromagnetic form factors are related to the dispersion integral of the spectral functions over $t \equiv -Q^{2}$, which depend on a number of empirical parameterizations, such as those for the higher-mass states and the time-like pion electromagnetic form factor. With each input value of charge and magnetic radii, the model can generate a unique set of $G_{E}^{p}$ and $G_{M}^{p}$ up to $Q^{2} \sim 1~\rm{GeV}^{2}$, with well-controlled theoretical uncertainty. This model was utilized to fit the Mainz 2010 cross-section data~\cite{A1:2010nsl, A1:2013fsc}, with the proton charge and magnetic radii as free parameters. Floating normalization parameters were also included in the fit, similar to what was done in the original Mainz analysis~\cite{A1:2010nsl, A1:2013fsc}. Since the DI$\chi$EFT model can predict the $Q^{2}$-dependency of $G_{E}^{p}$ and $G_{M}^{p}$, it naturally avoids the uncertainties in these floating parameters due to different choices of empirical fitters. Applying this model to the Mainz data with $Q^{2}$ up to 0.5~GeV$^{2}$ yields:
\begin{equation}
  \begin{split}
  &r_{E}^{p} = 0.842 \pm 0.002_{\rm{fit}} \pm 0.010_{\rm{theory}}~\rm{fm},  \\
  &r_{M}^{p} = 0.850 \pm 0.001_{\rm{fit}} \pm 0.010_{\rm{theory}}~\rm{fm},
  \end{split}
  \label{eq:alarcon_result}
\end{equation}
with a reduced $\chi^{2}$ of 1.39. The stability and robustness of this model were demonstrated by obtaining consistent results when extending the maximum $Q^{2}$ to 1 GeV$^{2}$ for the same dataset, using a rebinned version of the Mainz data~\cite{Lee:2015jqa}, and including the PRad data as well. Additionally, Horbatsch also employed DI$\chi$EFT to analyze the PRad and the Mainz ISR data~\cite{Horbatsch:2019wdn}, and found good agreement with parameterizations based on the muonic spectroscopic $r_{E}^{p}$ result and the fourth moment predicted by DI$\chi$EFT~\cite{Alarcon:2018irp}.

Lin, Hammer, and Mei{\ss}ner have conducted a series of studies applying dispersion theory to world data from space-like measurements, including unpolarized and polarized $ep$ elastic scattering~\cite{Lin:2021umk, Lin:2021umz}. They have recently extended their study to include time-like measurements, such as $e^{+}e^{-}$ annihilation and the reverse process~\cite{Lin:2021xrc}. 
Combining both space-like and time-like measurements, their final result on proton electromagnetic radii are~\cite{Lin:2021xrc}
\begin{equation}
  \begin{split}
  &r_{E}^{p} = 0.840 {^{+0.003}_{-0.002}}{^{+0.002}_{-0.002}}~\rm{fm},  \\
  &r_{M}^{p} = 0.849 {^{+0.003}_{-0.003}}{^{+0.001}_{-0.004}}~\rm{fm},
  \end{split}
  \label{eq:lin_result}
\end{equation}
where the first errors are statistical and the second errors are systematic uncertainties. In addition, they also determined the Zemach radius and the third Zemach moment, which are found to be consistent with Lamb shift and hyperfine splittings in $\mu$H measurement~\cite{Antognini:2013txn}. In fact, the dispersion analysis has consistently favored a smaller value of $r_{E}^{p}$~\cite{Hohler:1976ax, Mergell:1995bf, Hammer:2003ai, Belushkin:2006qa, Lorenz:2012tm, Lorenz:2014yda}, even before the proton charge radius puzzle arose. Interested readers can refer to the review on dispersion analysis and proton electromagnetic form factors~\cite{Pacetti:2014jai} for further information.

Gramolin and Russell~\cite{Gramolin:2021gln} proposed an alternative method to extract $r_{E}^{p}$ without relying on obtaining the slope of $G_{E}^{p}$ at $Q^{2} = 0$. Their approach involved parameterizing $F_1(Q^2)$ to directly relate to the even moments of the transverse charge density. By fitting these parameterizations to the measured cross-section data over a wide range of $Q^2$, they were able to determine the second moment of the transverse charge density, $\langle b^{2} \rangle$, through the fitted parameters. The proton charge radius $r_{E}^{p}$ is then obtained using the model-independent relation given by Eq.~\ref{eq:rp_b_relation}. The authors applied this method to the cross-section data from the Mainz 2010 experiment~\cite{A1:2010nsl} and found the proton charge radius to be
\begin{equation}
	r_{E}^{p} = 0.889(5)_{\rm{stat.}}(5)_{\rm{syst.}}(4)_{\rm{model}}~\rm{fm,}
\label{eq:Gramolin_result}
\end{equation}
which is consistent with the original Mainz 2010 results~\cite{A1:2010nsl, A1:2013fsc}. However, this study has been recently questioned by Boone~\textit{et al.}~\cite{Boone:2023ryu}. They applied the same approach to fit four different combinations of Mainz 2010 and PRad data: ``original Mainz'', ``re-binned Mainz'', ``original Mainz and PRad'', and ``re-binned Mainz and PRad''. While they were able to reproduce the result in \cite{Gramolin:2021gln} using the ``original Mainz'' data set, they found significantly different values of $r_{E}^{p}$ when fitting the other three combinations. Furthermore, Boone~\textit{et al.}~\cite{Boone:2023ryu} pointed out that the form factor parameterization proposed in \cite{Gramolin:2021gln} is unable to describe the form factor ratio data obtained with the polarization technique~\cite{Zhan:2011ji, Paolone:2010qc, Crawford:2006rz, Punjabi:2005wq} when $Q^{2} > 0.8$~GeV$^{2}$.

Atac \textit{et al.}~\cite{Atac:2020hdq} determined the second moment of the transverse charge density by fitting the slope of flavor-dependent Dirac form factors at $Q^2 = 0$ (Eq.~\ref{eq:transverse_radius_slope}). They performed a flavor decomposition of the proton and neutron form factor world data, assuming charge symmetry for the Dirac form factors. Various functional forms were used to fit $F_1^{u(d)}$ simultaneously and obtain the transverse mean-square radii $\langle b^{2}_{u(d)} \rangle$. The proton and neutron charge radii were then extracted from $\langle b^{2}_{u(d)} \rangle$. Their analysis yielded a proton charge radius of
\begin{equation}
	r_{E}^{p} = 0.852 \pm 0.002_{\rm{stat.}} \pm 0.009_{\rm{syst.}}~\rm{fm}.
\label{eq:Atac_result}
\end{equation}
The same procedure was also tested with the data set excluding the PRad data, which gave a result consistent with the previous one, but with a larger uncertainty, amounting to $0.857(13)$~fm. This study also reported the first extraction of neutron charge radius based on form factor data, with a value of $\langle {r_{E}^{n}}^{2} \rangle = -0.122 \pm 0.004_{\rm{stat.}} \pm 0.010_{\rm{syst.}}$~fm$^{2}$.

On the other hand, the extraction of charge radii from empirical fits and statistical methods remains an active research topic in the field. Hayward and Griffioen~\cite{Hayward:2018qij} extracted proton and deuteron charge radii by employing various empirical fits to the low-$Q^{2}$ $ep$ and $ed$ scattering data, prior to or including the Mainz 2010 data~\cite{A1:2010nsl}. The authors examined point-to-point uncertainties in these data sets and developed a comprehensive algorithm to investigate potential systematic biases from different fit functions. They also minimized the total uncertainty by selecting the optimal maximum $Q^{2}$ value for the fitted data set. The extracted $r_{E}^{p}$ is $0.842(4)$ fm, and it is dominated by the Mainz 2010 data set.
Barcus, Higinbotham, and McClellan\cite{Barcus:2019skg} reanalyzed the Mainz 2010 data set using a similar polynomial expansion (Eq.~\ref{eq:poly_fitter}) as in the original Mainz 2010 analysis~\cite{A1:2010nsl, A1:2013fsc}. The authors were able to reproduce the Mainz results when using an unbounded polynomial fit, similar to what had been done in the original analysis. However, by imposing an additional constraint on the fitting parameters in the polynomial such that they have successively alternating signs, which makes the polynomial approximately completely monotonic, they obtained a much smaller $r_{E}^{p}$ of 0.854~fm.
Paz~\cite{Paz:2020prs} analyzed the PRad data using a high-order $z$-expansion (Eq.~\ref{eq:z_expansion_fitter}), but with bounded coefficients in the fitter. This approach helps to prevent underestimating the $r_{E}^{p}$ uncertainty due to truncation of the expansion, while avoiding the growth of uncertainty from unbounded high-order terms in the $z$-expansion. The author found that the extracted $r_{E}^{p}$ stabilized quickly after the third power $z$-expansion, but the statistical uncertainty is approximately 50\% larger compared to the published result of the PRad experiment~\cite{Xiong:2019umf}.

Zhou \textit{et al.}~\cite{Zhou:2021gyh} proposed a possible explanation for the $G_{E}^{p}$ discrepancy between the PRad and Mainz 2010 data within the $Q^{2}$ range of 0.01GeV$^{2}$ to 0.06GeV$^{2}$. The authors observed that the Rational (1, 1) functional form (Eq.~\ref{eq:rational_fitter}) provides an excellent approximation to the state-of-the-art DI$\chi$EFT model~\cite{Alarcon:2020kcz, Alarcon:2018zbz, Alarcon:2018irp, Alarcon:2017lhg}. They then attempted to parameterize both $G_{E}^{p}$ and $G_{M}^{p}$ using the Rational (1, 1) function and fit them simultaneously to the Mainz 2010 cross-section data up to $Q^{2} = 0.5$~GeV$^{2}$. This approach mostly removed the discrepancy of $G_{E}^{p}$ between PRad and Mainz 2010, suggesting that the discrepancy was mainly due to underestimated systematic uncertainties associated with determining the normalization factors.

Borah, Hill, Lee, and Tomalak~\cite{Borah:2020gte} conducted a thorough re-analysis of world data, including $ep$ and $en$ elastic scattering data, as well as form factor ratio measurements of polarized $ep$ scattering. In the main analysis, the PRad data were excluded due to the unavailability of uncertainty correlations. The authors applied a more robust radiative correction to earlier unpolarized elastic scattering data, which included hadronic vertex, hadronic vacuum polarization, and TPE. The study obtained a compact representation of the nucleon form factors, with the proton and neutron charge radii fixed by high-precision external constraints from the $\mu$H spectroscopy and neutron scattering length measurements, respectively. In the appendix, the PRad data were re-analyzed without the external constraint from the $\mu$H measurement. By using a third-order $z$-expansion, $r_{E}^{p}$ was determined as $0.836(19)$~fm, with the PRad statistical and systematic uncertainties added in quadrature. When combined with the Mainz data set, an eighth-order $z$-expansion yielded $r_{E}^{p}$ of $0.843(11)$~fm.

Cui \textit{et al.}~\cite{Cui:2021vgm} presented an innovative approach to extracting the proton charge radius using the statistical Schlessinger Point Method (SPM). Unlike traditional empirical fits, this method does not rely on a particular functional form and captures both local and global features of the underlying curve using a set of continuous fraction interpolations. Applying the SPM method to the PRad data, they obtained $r_{E}^{p} = 0.838 \pm 0.005_{\rm{stat.}}$~fm, with the statistical uncertainty estimated with the bootstrap method. The same approach was applied to the Mainz 2010 data set in the $3.8\times10^{-3}< Q^{2}/{\rm{GeV}^{2}} < 1.4\times10^{-2}$ range, yielding $r_{E}^{p} = 0.856 \pm 0.014_{\rm{stat.}}$~fm. Including all data from Mainz 2010 experiment resulted in almost the same central value, but with a slightly larger statistical uncertainty. Combining both PRad and Mainz 2010 data, the final result was
\begin{equation}
	r_{E}^{p} = 0.847 \pm 0.008_{\rm{stat.}}~\rm{fm}.
\label{eq:Cui_result}
\end{equation}
The SPM method was also applied to the Mainz 2010 data set to extract the proton magnetic radius~\cite{Cui:2022fyr, Cui:2021skn}, which was found to be $r_{M}^{p} = 0.817 \pm 0.027_{\rm{stat.}}$~fm.

\subsection{Progress from Lattice QCD}
Lattice QCD has proven to be a powerful tool for providing \textit{ab initio} calculations of various hadronic observables, including the nucleon charge, spin, and parton distribution functions~\cite{Ma:2014jla, Musch:2010ka, Bhattacharya:2016zcn}. The low $Q^{2}$ nucleon form factor and the proton charge radius have drawn attention from the Lattice QCD community~\cite{Djukanovic:2021qxp} since the discovery of the proton charge radius puzzle.  To investigate these topics, in principle one needs to calculate both the isovector ($G^{v} \equiv G^{u-d}$) and isoscalar ($G^{s} \equiv G^{u+d}$) nucleon form factors. These two quantities are related to the difference and sum of the proton and neutron form factors, respectively.

\begin{figure}
     \centering
     \begin{subfigure}[b]{0.45\textwidth}
         \centering
         \includegraphics[width=\textwidth]{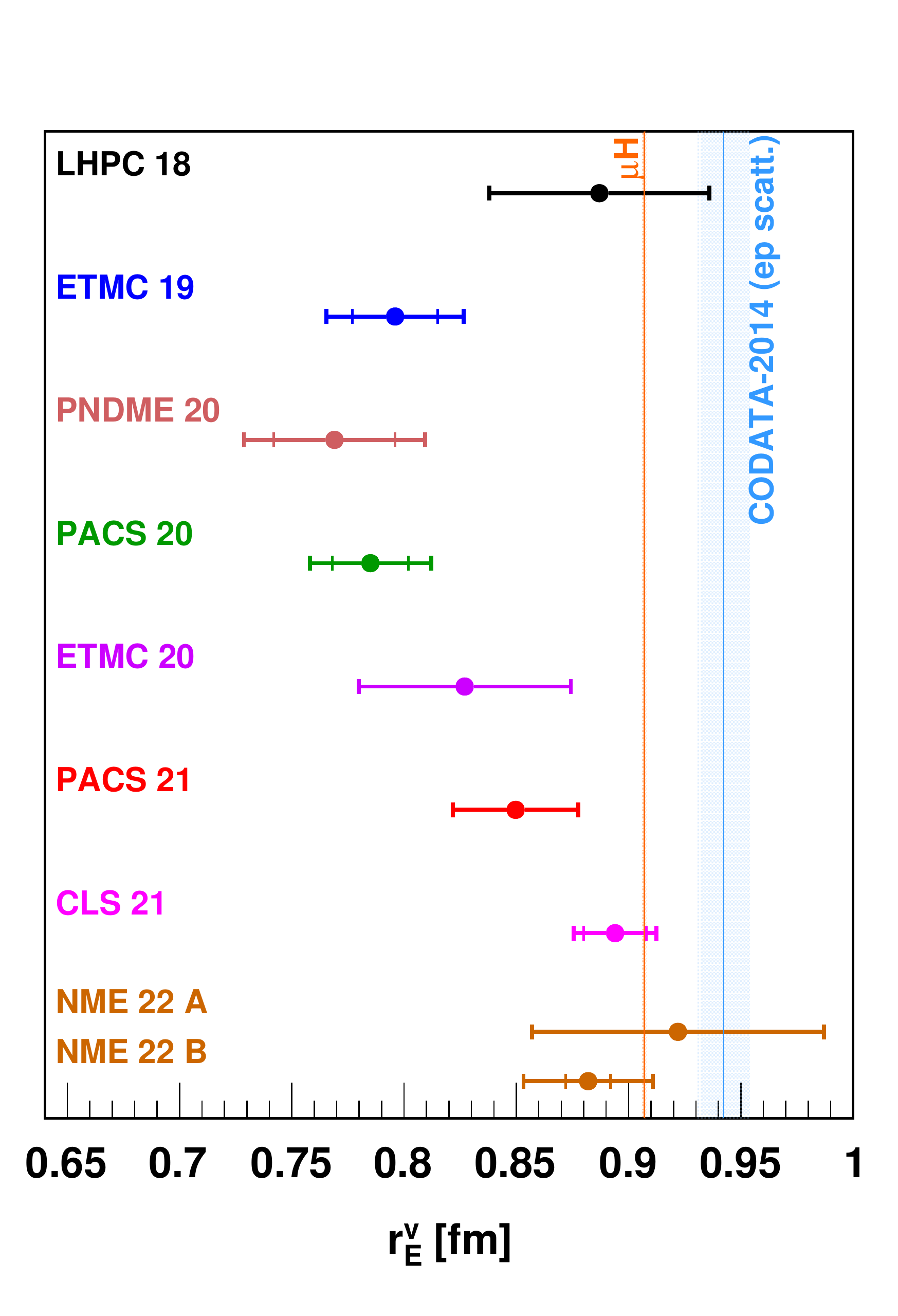}
     \end{subfigure}
     \hfill
     \begin{subfigure}[b]{0.45\textwidth}
         \centering
         \includegraphics[width=\textwidth]{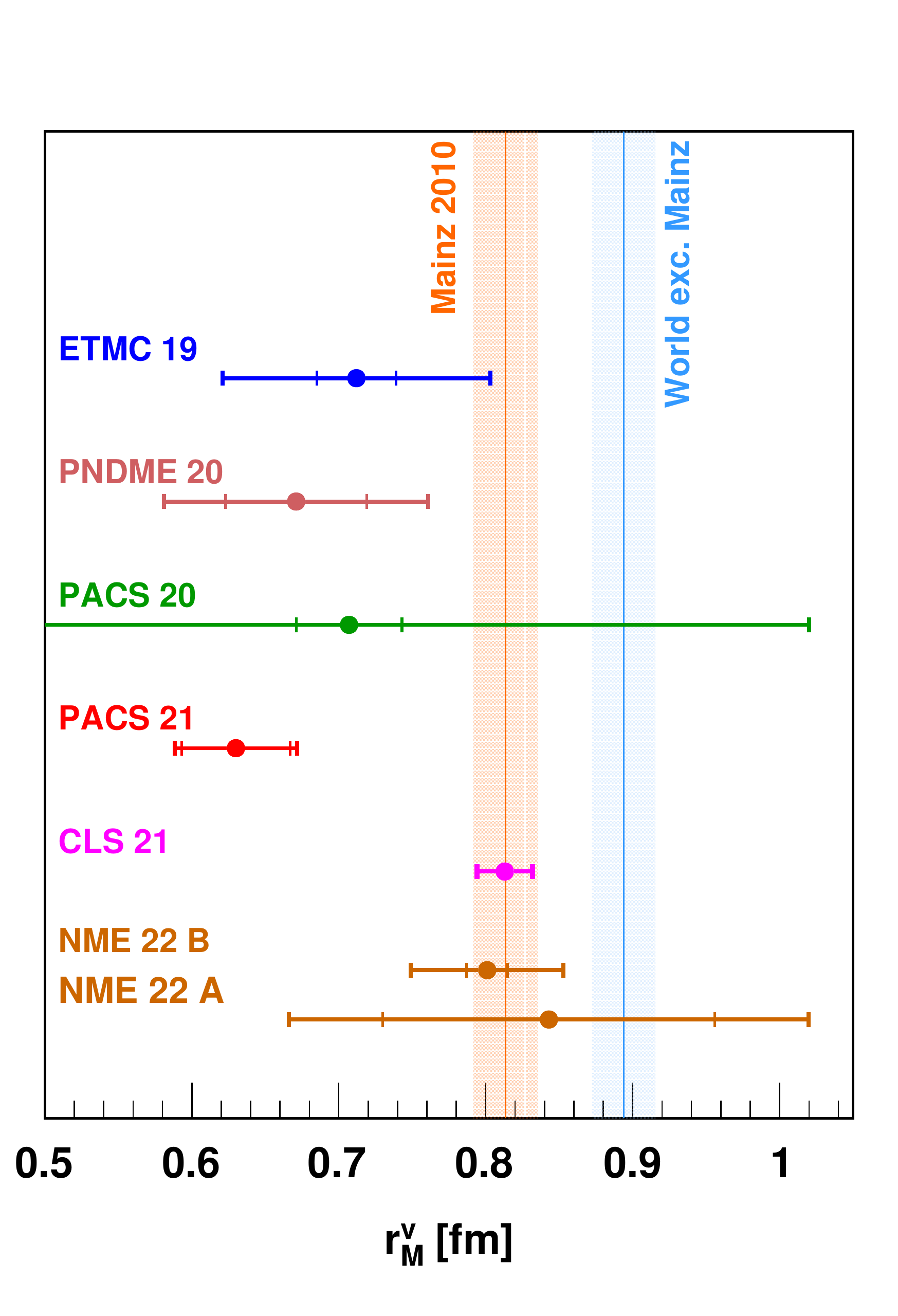}
     \end{subfigure}
     \hfill
        \caption{\textbf{Left panel}: the isovector charge radius from recent Lattice QCD calculations~\cite{Hasan:2017wwt, Alexandrou:2018sjm, Jang:2019jkn, Shintani:2018ozy, Alexandrou:2020aja, Ishikawa:2021eut, Park:2021ypf, Djukanovic:2021cgp}. The vertical line and band in orange color are obtained based on the $r_{E}^{p}$ result from the muonic hydrogen spectroscopic experiment~\cite{Antognini:2013txn}, and the blue vertical line and band are obtained based on the CODATA-2014 $r_{E}^{p}$ compilation~\cite{Mohr:2015ccw}. Information about the neutron charge radius is obtained from Particle Data Group (PDG)~\cite{ParticleDataGroup:2020ssz}. \textbf{Right panel}: the isovector magnetic radius from recent Lattice QCD calculations. The vertical line and band in orange color are obtained based on the $r_{M}^{p}$ result from Lee \textit{et al.}~\cite{Lee:2015jqa} for the Mainz 2010 data set, and the blue line and band are obtained based on the result from the same analysis, for the world data excluding the Mainz 2010 data set. Information about the neutron magnetic radius is obtained from PDG~\cite{ParticleDataGroup:2020ssz}.}
        \label{fig:lattice_radius}
\end{figure}

\begin{figure}[H]
\centering
\includegraphics[width=12 cm]{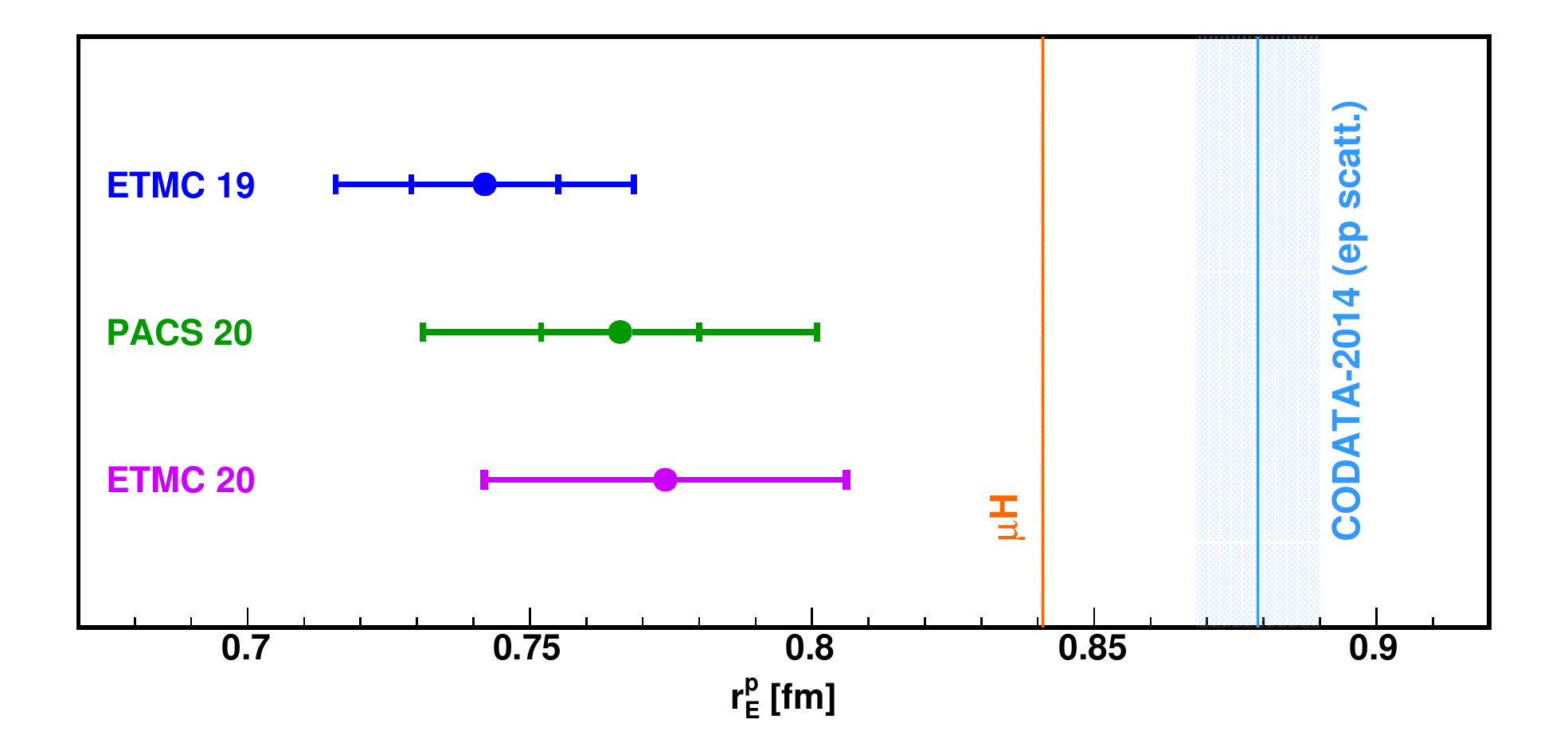}
\caption{The proton charge radius $r_{E}^{p}$ from recent lattice QCD calculations~\cite{Alexandrou:2018sjm, Shintani:2018ozy, Alexandrou:2020aja}, and compared to the $\mu$H measurement~\cite{Antognini:2013txn} and the CODATA-2014 recommended value~\cite{Mohr:2015ccw}. ETMC 19 calculation~\cite{Alexandrou:2018sjm} includes disconnected contributions.}
\label{fig:lattice_charge_radius}
\end{figure}

The isovector form factors only require consideration of connected diagrams, while for isoscalar form factors, the much more complicated and computationally expensive disconnected diagrams must be dealt with. Nucleon electromagnetic form factors are a linear combination of isovector and isoscalar form factors. The isovector and isoscalar radii can also be extracted by taking the corresponding form factor slope at $Q^{2} = 0$. In many cases, the isovector and isoscalar radii need to be extracted from empirical fits to the lattice form factor data, as analogous to the extraction of charge radius from experimental data. Recently, certain methods have been developed to implement derivatives on the correlator level and thus allow direct computation of the form factor slope or nucleon radii~\cite{Hasan:2017wwt, Alexandrou:2020aja, Ishikawa:2021eut}. Such approaches avoid the systematic biases associated with form factor fitting and extrapolation to $Q^{2} = 0$. Some of the recent calculations for the proton isovector electromagnetic radii are shown in Fig.~\ref{fig:lattice_radius}, and the proton charge radius results are shown in Fig.~\ref{fig:lattice_charge_radius}. Despite difficulties such as finite lattice size and contamination from excited states, the lattice QCD community continues to make progress in reducing uncertainties of both form factors and radii. However, higher precision calculations are still needed to have a significant impact on the proton charge radius puzzle.

\unskip
The \textit{ab initio} calculation of proton electric form factor at low $Q^2$ is also of great interest, particularly in the $Q^2$ range between 0.01 to 0.06 GeV$^2$, due to the large discrepancy observed between the PRad data~\cite{Xiong:2019umf} and the Mainz 2010 data~\cite{A1:2010nsl} (see Fig.~\ref{fig:recent_ff}). Although lattice calculations in the very low $Q^2$ region are challenging due to the need for a large lattice setup, high-precision lattice calculations in this range are very interesting and can provide important input to this data tension. Further information on recent Lattice QCD progress regarding nucleon form factors and radii can be found in the recent proceeding by Djukanovic~\cite{Djukanovic:2021qxp}.


\unskip

\section{Remaining Issues in Lepton Scattering Experiments and Possible Explanation}
\label{sec:remaining_issue}

Over the past decade, numerous experimental efforts have shed light on the proton charge radius puzzle through hydrogen spectroscopy, with many favoring smaller values of $r_{E}^{p}$, including the recent high-precision ordinary hydrogen spectroscopic results from Beyer \textit{et al.}~\cite{Beyer:2017gug}, Grinin \textit{et al.}~\cite{Grinin2020} and Bezginov \textit{et al.}~\cite{Bezginov:2019mdi}. In particular, the last one measured the 2S$_{1/2}$-2P$_{1/2}$ transition frequency, which strongly supports the smaller radius due to not requiring precise knowledge of the Rydberg constant in such Lamb shift measurements. In addition, the recent review by Peset, Pineda, and Tomalak~\cite{Peset:2021iul} has clarified the consistency of $r_{E}^{p}$ definition between muonic and ordinary hydrogen spectroscopic measurements. As a result, the community can no longer ignore the high-precision $\mu$H results, and both Particle Data Group and CODATA have incorporated them in their latest publications~\cite{ParticleDataGroup:2022pth, Tiesinga:2021myr}, yielding $r_{E}^{p} = 0.8409(4)$~fm and $r_{E}^{p} = 0.8414(19)$~fm, respectively. It is widely accepted that the original 7$\sigma$ discrepancy in the proton charge radius puzzle has been partially resolved. However, several issues remain, including the tension between the latest high-precision hydrogen spectroscopic measurement~\cite{Brandt:2021yor} and the $\mu$H results, which requires further investigation and clarification.

The issues faced by the lepton scattering community are perhaps more severe. To date, the PRad experiment \cite{Xiong:2019umf} is the only new unpolarized lepton scattering experiment with sufficient precision to impact the puzzle since 2010. While its result is consistent with the $\mu$H measurements, the disagreement between PRad and other modern scattering experiments calls for further experimental efforts to cross-check the systematic uncertainties. Moreover, in addition to the $r_{E}^{p}$ puzzle, the observed $G_{E}^{p}$ data tension between the Mainz 2010 and PRad experiments needs to be investigated and understood. This discrepancy is already present in the lower $Q^{2}$ region and becomes more prominent at higher $Q^{2}$ values between 0.01GeV$^{2}$ and 0.06GeV$^{2}$. Several possible factors could explain this data tension, particularly those related to the kinematic factor $\epsilon$, which is quite different between PRad and previous experiments.

The radiative correction could possibly contribute to the discrepancy, for which PRad and Mainz 2010 experiments used different recipes for internal radiative corrections. PRad used first-order calculations beyond the URA from Akushevich \cite{Akushevich:2015toa}, while Mainz 2010 followed the recipes from Maximon-Tjon \cite{Maximon:2000hm} and Vanderhaeghen \cite{Vanderhaeghen:2000ws}. Both experiments relied on full simulations of the radiative effects to unfold the Born-level cross-sections, and they had achieved sub-percent agreement between the simulation and data \cite{Xiong:2020kds, A1:2013fsc}. However, PRad found that the next-to-next leading order (NNLO) contributions for the M{\o}ller process, which were roughly estimated in their analysis, could significantly contribute to the systematic uncertainty in determining the form factor slope because of kinematic dependent corrections. This finding motivated improved calculations of QED radiative effects for the future PRad-II experiment~\cite{Afanasev:2020hwg}. Moreover, the two experiments might have very different TPE corrections, given their different $Q^{2}$ and $\epsilon$. Tomalak and Vanderhaeghen \cite{Tomalak:2018ere, Tomalak:2015aoa, Tomalak:2014sva} have shown that the TPE correction is relatively easier to determine in the low-$Q^{2}$ and forward angular region and contributes no more than 0.2\% to the cross section in the PRad kinematic coverage, which is a negligible contributing factor comparing to the total uncertainties. However, for the Mainz data set, this correction is expected to be more significant due to the $Q^{2}$ and $\epsilon$ ranges of their dataset \cite{A1:2013fsc, Lorenz:2014yda, Tomalak:2017shs, Tomalak:2018ere}.

The separation of the proton's electric and magnetic form factors can also impact the extracted charge radius. Given that the measured $ep$ elastic cross section includes contributions from both $G_{E}^{p}$ and $G_{M}^{p}$, systematic uncertainties associated with $G_M^p$ and $G_E^p$ are often correlated, meaning that an underestimated $G_{E}^{p}$ can lead to an overestimated $G_{M}^{p}$ at the same kinematic point, and vice versa. One less well-known discrepancy in this area pertains to the proton's magnetic form factor and its corresponding radius $r_{M}^{p}$. Specifically, as noted in the original Mainz 2010 analysis~\cite{A1:2010nsl} and subsequently by Lee \textit{et al.}~\cite{Lee:2015jqa}, a 2.7$\sigma$ difference was observed in $r_{M}^{p}$ between the Mainz 2010 data and other world data.

Last but not least, the observed discrepancy may be magnified by underestimated or overlooked systematic uncertainties, such as those arising from the fitting procedure and those associated with the experimental apparatus. Recent re-analyses~\cite{Barcus:2019skg, Zhou:2021gyh} have shown that fitting the Mainz 2010 data with a more constrained functional form leads to a different set of normalization parameters and $G_E^{p}$ that are much more consistent with the PRad results. Additionally, the PRad experiment employed a hybrid calorimeter as its primary detector for particle energies, with outer lead-glass modules having much worse resolution compared to the inner PbWO$_{4}$ modules. The kinematic range covered by the lead-glass modules largely overlaps with the $Q^2$ range where the discrepancy is most prominent, highlighting the need for further investigations into the corresponding systematic uncertainties. Fortunately, upcoming experiments, which we will discuss in the following section, are planning to use advanced target and detector systems, as well as state-of-the-art analysis methods, to enable better control over these types of systematics.

\section{Future Lepton-proton Scattering Experiments}

\begin{figure}[H]
\centering
\includegraphics[width=12 cm]{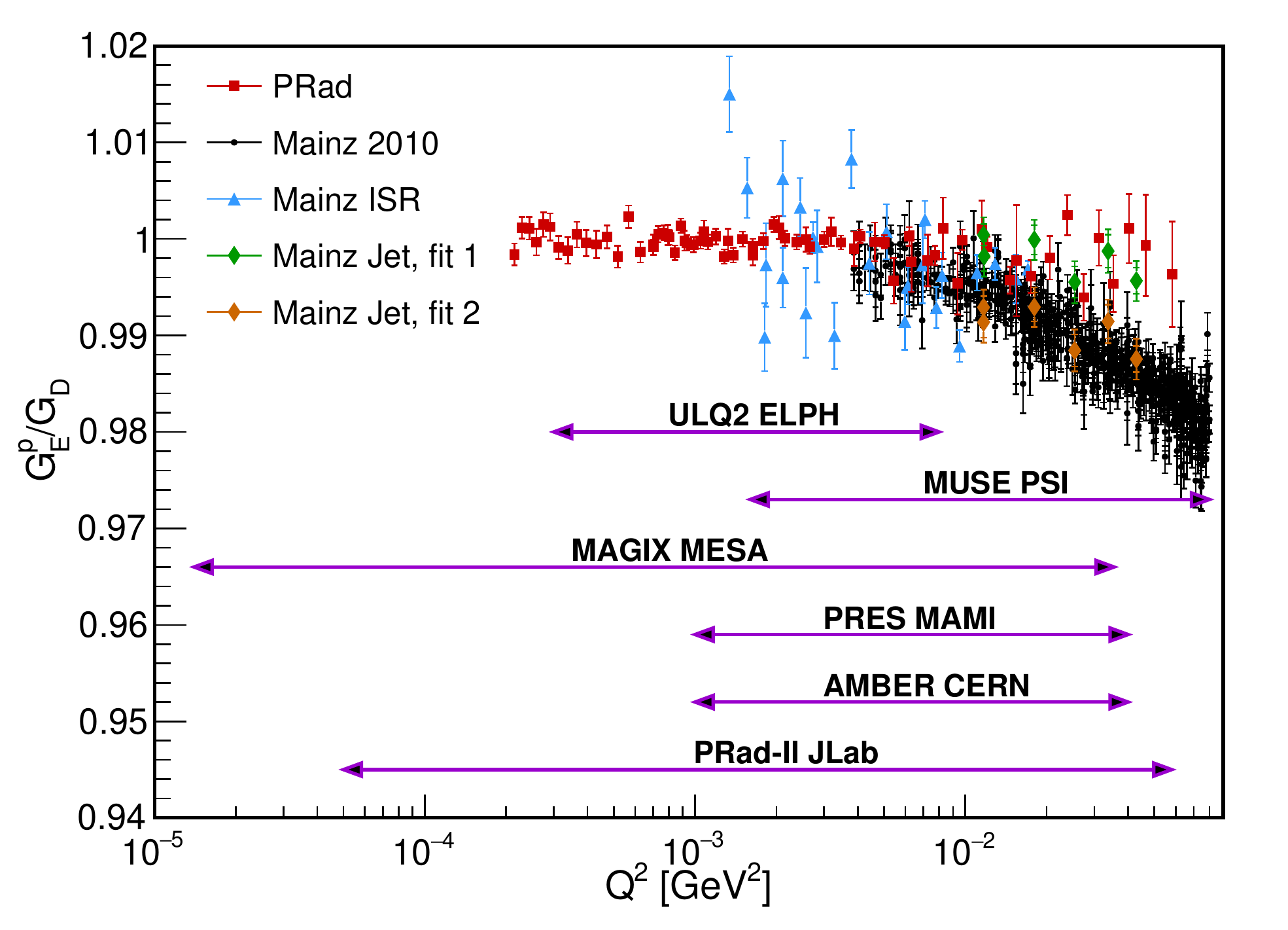}
\caption{The proton electric form factor, normalized by the standard dipole form factor, from the PRad experiment~\cite{Xiong:2019umf} (red squares), the Mainz 2010 experiment~\cite{A1:2010nsl} (black dots), the Mainz ISR experiment~\cite{Mihovilovic:2019jiz} (blue triangles), and Mainz jet target experiment~\cite{A1:2022wzx}, where green diamonds are fitted to PRad parameterization and orange diamonds are fitted to Mainz 2010 parameterization. Also shown on top, are the expected $Q^{2}$ coverages from future experiments including MUSE~\cite{Kohl:2022lex, Cline:2021ehf, MUSE:2017dod}, PRad-II~\cite{PRad:2020oor}, AMBER~\cite{Quintans:2022utc, COMPASSAMBERworkinggroup:2019amp}, PRES~\cite{Belostotski:2019qum, Vorobyev:2019hpy}, MAGIX~\cite{A1:2021njh} and ULQ2 experiments~\cite{ULQ2, Suda:2022hsm}.}
\label{fig:future_experiments}
\end{figure}

Several new lepton-proton elastic scattering experiments are currently being prepared or are taking data, motivated by the unresolved proton charge radius puzzle and the data tension in the form factor measurements. Most of these experiments aim to cover the low $Q^{2}$ region, where the PRad data and the Mainz 2010 data exhibit a large discrepancy (see Fig.~\ref{fig:future_experiments}). These experiments employ vastly different experimental techniques and each possesses its own unique advancements. They are expected to provide diverse inputs to address the outstanding issues in the field. In the following subsections, we will introduce the experimental designs and special features of these next-generation experiments that focus on low-$Q^{2}$ lepton-proton elastic scattering. For further details, readers can refer to recent reviews by Gao and Vanderhaeghen\cite{Gao:2021sml}, as well as Karr, Marchand, and Voutier~\cite{Karr:2020wgh}.

\unskip

\subsection{MUSE Experiment}
Up to this point, all measurements of the proton charge radius using the lepton scattering technique have been obtained with electrons as the probe. The first measurement of $r_{E}^{p}$ using muons is expected to come from the MUon Scattering Experiment (MUSE) \cite{MUSE:2017dod, Cline:2021ehf}, which is currently collecting data at the Paul Scherrer Institut (PSI). This experiment aims to measure lepton-proton elastic scattering cross sections using both electrons and muons, as well as positrons and antimuons, with data from the same beam polarity being collected simultaneously. This unique experiment will provide valuable insights into the proton charge radius puzzle. Firstly, a comparison between electronic and muonic measurements will be a direct test for lepton-universality violation and any related new physics. Secondly, this comparison can test our understanding of radiative corrections (RC). Muons have nearly 200 times the mass of electrons and thus have much smaller radiative effects. The possible observation of a systematic difference between electronic and muonic measurements would indicate issues with RC, which is one of the main suspects for the discrepancy between the PRad data and earlier measurements. Furthermore, the use of both positive and negative polarities of the incoming lepton beam allows control of the contribution from the two-photon exchange (TPE) diagrams \cite{Ahmed:2020uso, Tomalak:2018ere, Tomalak:2017shs, Tomalak:2017owk, Tomalak:2016vbf, Tomalak:2015aoa, Tomalak:2014sva}. Depending on the kinematics, TPE diagrams contribute to the cross section by up to 1\% \cite{Cline:2021ehf, Tomalak:2018jak, Tomalak:2015hva, Tomalak:2014dja}, and this contribution is notoriously difficult to precisely calculate. However, taking the difference between cross-sections with different beam polarities can effectively constrain this contribution.

\begin{figure}[H]
\centering
\includegraphics[width=10 cm]{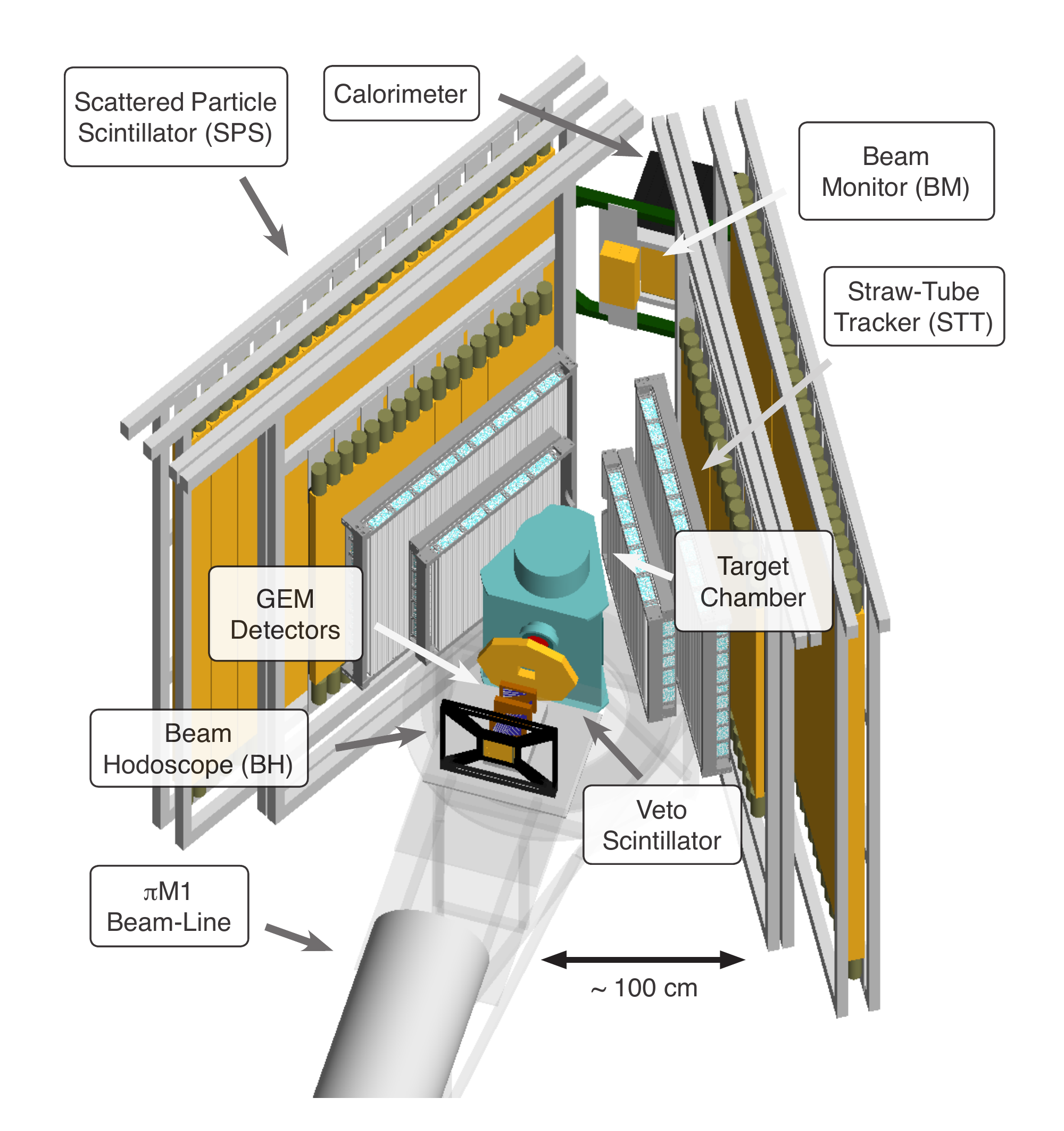}
\caption{The experimental apparatus of the MUSE experiment. Figure credit: Steffen Strauch.}
\label{fig:MUSE_setup}
\end{figure}

The MUSE experiment utilizes the $\pi$M1 beamline at PSI to measure the cross sections for lepton-proton elastic scattering at three beam energies: 115, 161, and 210 MeV. Since this beamline delivers a mixture of pions, muons, and electrons, the experiment requires excellent particle identification (PID) among these species of incident particles. Moreover, several beam properties, such as emittance and momentum bite, are not as good as those of modern primary electron beams~\cite{Cline:2021vlw}. To overcome these challenges, the experiment uses a beam hodoscope (shown in Fig.~\ref{fig:MUSE_setup}) to obtain precise timing measurements as the incident particles pass through. The time-of-flight obtained by combining these measurements with information from the accelerator RF provides a precise PID. Three GEM detectors located immediately after the hodoscope determine the incident angles with high resolution and can achieve angular resolutions at the level of 1 mrad. Between the GEMs and the target chamber, a veto scintillator detector reduces the background and rejects particles that have decayed in flight. The target chamber contains three targets: a liquid hydrogen target for physics production, an empty target for background studies, and a carbon target for detector alignments. Scattered leptons exit the target chamber through the side walls, with their scattering angles measured by the straw-tube trackers and timing measured by the scattered particle scintillators. Meanwhile, un-scattered leptons reach the downstream beam monitor and eventually the calorimeter. The former provides flux and timing measurements, and the latter can be used to control radiative effects, particularly those from initial state radiation (as shown in the left panel of Fig.~\ref{fig:BH_diagram}).

The MUSE experiment's detector setup provides a scattering-angle coverage from 20$^{\circ}$ to 100$^{\circ}$, with an azimuthal-angle coverage of approximately 30\% of 2$\pi$. Together with the three beam energies mentioned above, the apparatus enables the experimental study of lepton-proton elastic scattering over a $Q^{2}$ range approximately from 0.0016 to 0.08 GeV$^{2}$. The expected statistical uncertainty for the cross-section measurements is better than 1\%, and the uncertainty on $r_{E}^{p}$ is expected to be around $0.01$ fm for all four different incident particle species. The MUSE experiment is presently acquiring data and is anticipated to achieve the required statistics within the next two years~\cite{Kohl:2022lex}. The experiment's results are highly anticipated due to its numerous unique features.

\unskip

\subsection{PRad-II Experiment}
The PRad-II experiment~\cite{PRad:2020oor} has been approved by the Jefferson Lab program advisory committee (PAC) with the highest scientific rating. Building on the success of the original PRad experiment, PRad-II will retain the advantages of its non-magnetic calorimetric configuration and windowless target, while also improving its experimental apparatus and analysis methods to reduce the total uncertainties of $r_{E}^{p}$ and $G_{E}^{p}$ by a factor of about 4. This will enable an unprecedented precision of $r_{E}^{p}$ (0.0036 fm) from scattering experiments and help to address the $G_{E}^{p}$ data tension between PRad and Mainz 2010 measurements.

\begin{figure}[H]
\centering
\includegraphics[width=12 cm]{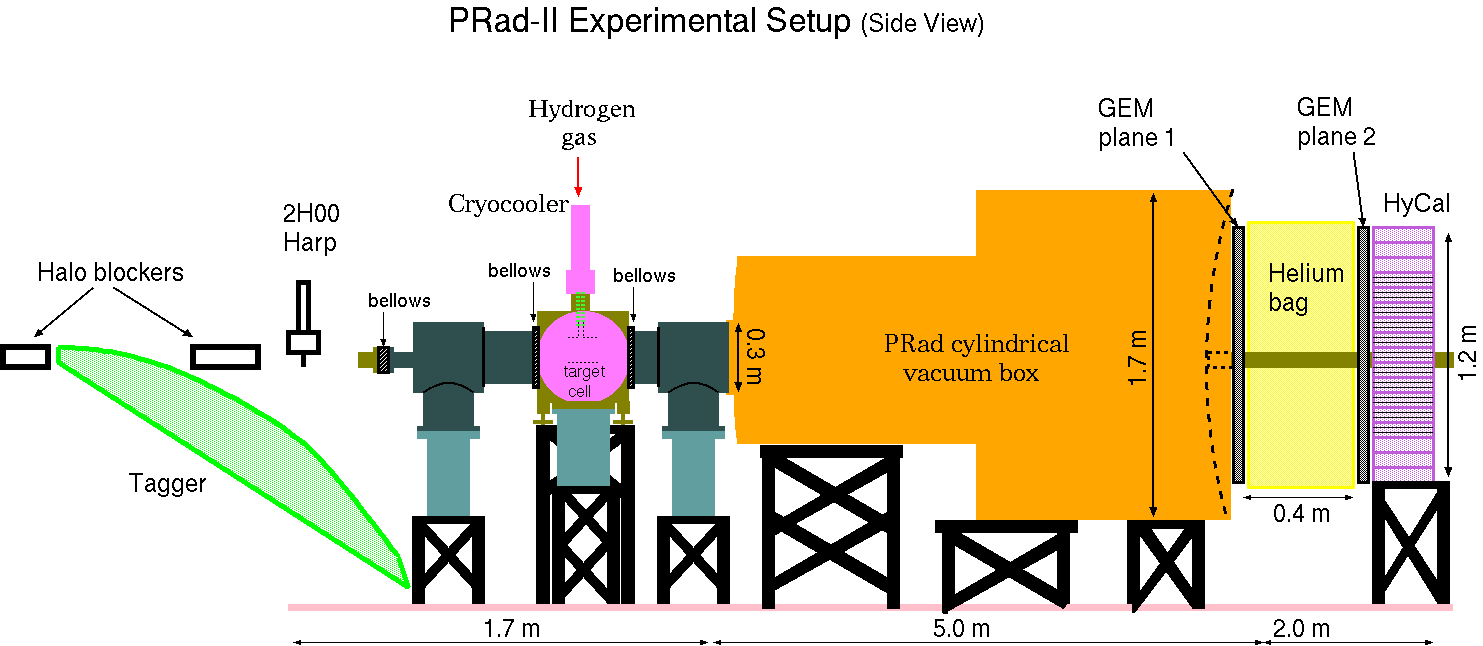}
\caption{The experimental apparatus of the PRad-II experiment. Figure credit: A. Gasparian \textit{et al.} \cite{PRad:2020oor}}
\label{fig:PRad2_setup}
\end{figure}
\unskip

The experimental setup for PRad-II is depicted in Fig.~\ref{fig:PRad2_setup}. A major improvement of this setup is the addition of a second GEM plane, which, when separated by 40 cm from the other GEM plane, provides sufficient leverage for a precise vertex-$z$ reconstruction that rejects the backgrounds from beam halo and residual gas. Additionally, the second GEM plane significantly improves the measurement of each GEM's efficiency as the other GEM and HyCal can both serve as reference detectors, allowing a coincidence cut to better reconstruct the events. During the PRad experiment, the dominant limiting factor for the GEM efficiency measurement was the positional resolution of HyCal, which solely served as the reference detector. The precise knowledge of GEM efficiency in PRad-II allows a global luminosity normalization with the integrated M{\o}ller cross-section over a broad $Q^2$ range, in which each $Q^2$ bin of the elastic $ep$ cross section needs a correction of the detector efficiencies. This normalization method, as compared to the bin-by-bin $ep$/$ee$ ratio normalization used in PRad, avoids introducing Q2-dependent systematic uncertainties from the calculations of M{\o}ller scatterings into the ep cross-section and provides an additional handle to study the systematic uncertainties associated with normalization.

Another major improvement is to replace all lead-glass modules with PbWO$_4$ modules, which results in over 2.5 times better energy resolution at large scattering angles (approximately $\theta > 3.8^{\circ}$). The lead-glass modules showed significant non-linearity responses for energy depositions of scattered electrons, leading to dominating systematic uncertainties for cross-section results in the high $Q^{2}$ region of PRad data. Given the fact that the $Q^2$ range of the observed discrepancy between PRad and Mainz 2010 data largely overlaps with the $Q^{2}$ region covered by the lead-glass modules, it is important to investigate this $Q^2$ range with the PbWO$_4$ calorimeter in the PRad-II experiment.

The PRad-II setup also includes a new scintillating detector mounted 25 cm downstream of the target cell to expand the detector acceptance for double-arm M{\o}ller events at very forward angles. In the original PRad experiment, the $ep$ and M{\o}ller events are distinguished by their energies with a scattering angle $\theta > 0.7^{\circ}$. However, the energy of the scattered electrons from these two processes converge at smaller scattering angles, making it impossible to differentiate them even with the high-resolution PbWO$_4$ calorimeter. While M{\o}ller events can still be identified if both electrons are detected, the electron with a larger scattering angle falls outside of the HyCal acceptance in this kinematic range. The new scintillating detector will be able to detect these missing M{\o}ller electrons, extending the lowest angular coverage down to 0.5$^{\circ}$, or approximately $5\times10^{-5}$ GeV$^{2}$ with the lowest beam energy. This allows PRad-II to reach an unprecedented low-$Q^2$ region for lepton-scattering measurements.

\begin{figure}
     \centering
     \begin{subfigure}[b]{0.54\textwidth}
         \centering
         \includegraphics[width=\textwidth]{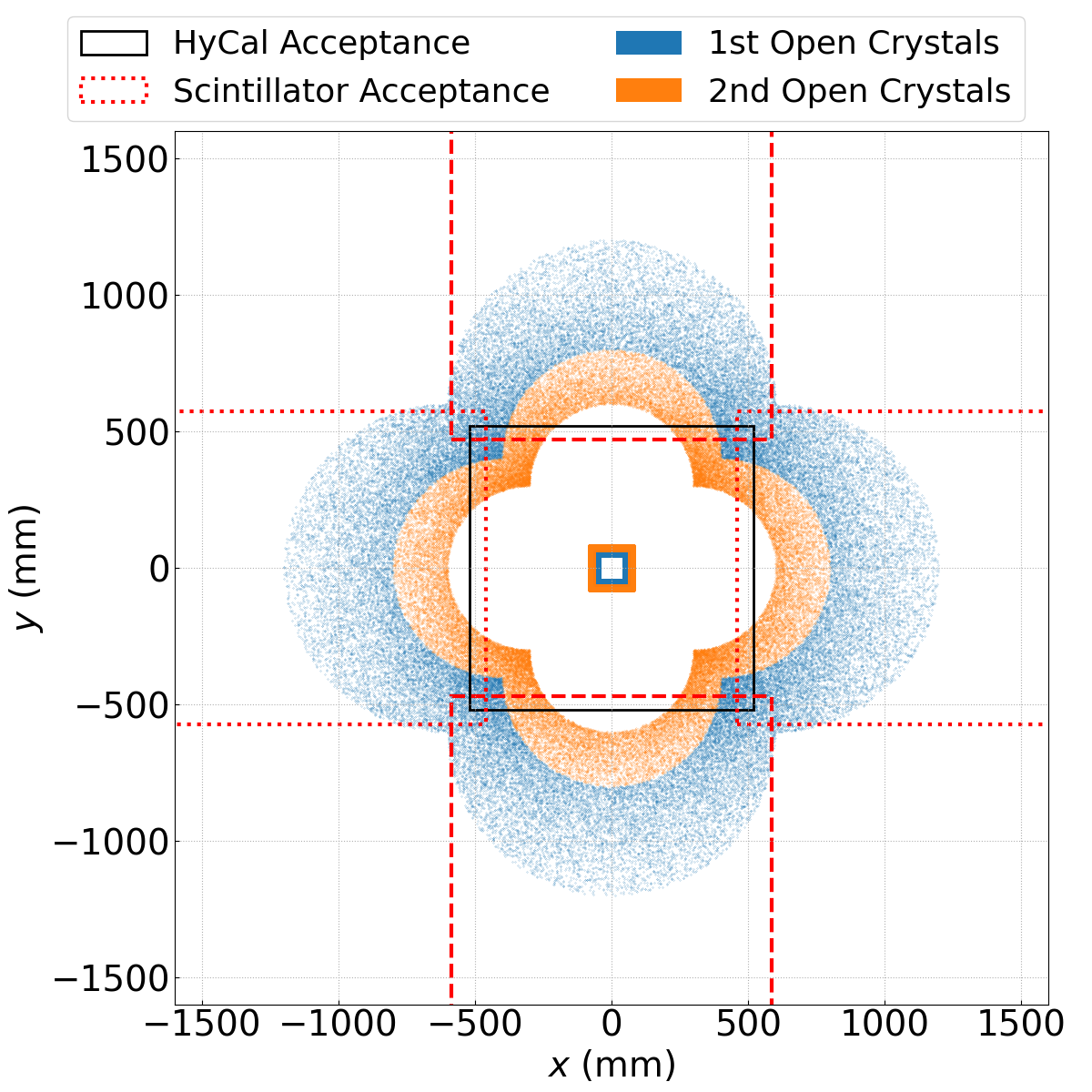}
     \end{subfigure}
     \hfill
     \begin{subfigure}[b]{0.45\textwidth}
         \centering
         \includegraphics[width=\textwidth]{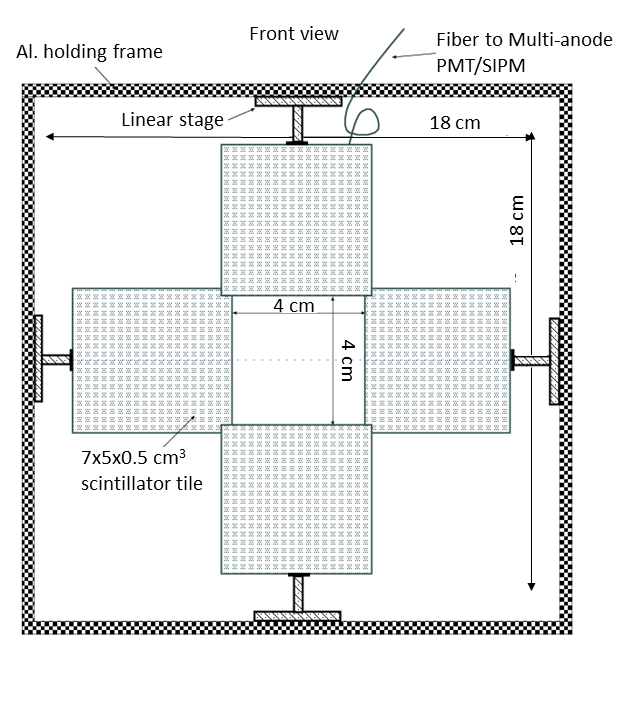}
     \end{subfigure}
     \hfill
        \caption{\textbf{Left panel}: The distribution of M{\o}ller electrons, when the other one is detected by the two opened innermost layers of HyCal. Red dash lines show the boundaries of the scintillator tiles. \textbf{Right panel}: A schematic view of the scintillator
detector to enhance detection of the M{\o}ller electrons. Figure credit: A.~Gasparian \textit{et al.}~\cite{PRad:2020oor}}
        \label{fig:PRad2_scintillating_detector}
\end{figure}
\unskip

In addition to the detector upgrades, the collaboration plans to replace the Fastbus readout system of the calorimeter with a full flash-ADC-based readout. This new system will significantly increase the event-rate capacity and enable an event-wise measurement of the electronic noise. It will also provide much better timing measurements to reject accidentals and improve trigger efficiency. Furthermore, the collaboration aims to improve the RC calculation by including NNLO diagrams for both $ep$ and M{\o}ller scatterings beyond the ultra-relativistic approximation, where the electron mass is not neglected~\cite{Afanasev:2020hwg}. This improvement is expected to significantly reduce the systematic uncertainty of $r_{E}^{p}$ from RC.

The PRad-II experiment plans to collect data using 0.7, 1.4, and 2.1 GeV electron beams, covering $Q^{2}$ values approximately from $5\times10^{-5}$ to 0.056 GeV$^{2}$. With approximately 24 days of production runs, the anticipated statistical uncertainty on $r_{E}^{p}$ will be about 0.0017 fm, which is 4.4 times smaller than that of PRad. Additionally, the collaboration has conducted a comprehensive analysis of the projected systematic uncertainty, taking into account all aspects based on experiences gained from the PRad experiment. When summed in quadrature, the total uncertainty on $r_{E}^{p}$ is expected to be about 0.0036 fm, nearly a factor-of-4 improvement over the PRad result.

\subsection{Compass++/AMBER Experiment}
The AMBER experiment~\cite{COMPASSAMBERworkinggroup:2019amp, Quintans:2022utc}, approved by CERN, will use a 100 GeV muon beam from the M2 beam line of the Super Proton Synchrotron (SPS) at CERN. By measuring the scattered muons in the extreme forward angular region (at the level of 1 mrad), the experiment can reach a very low $Q^{2}$ range ($0.001<Q^{2}<0.04$ GeV$^{2}$) that minimizes radiative effects and $G_{M}^{p}$ contribution. However, measuring $Q^{2}$ down to 0.001 GeV$^{2}$ in the AMBER setup requires a high scattering-angle resolution of below 100~${\mu}$rad. The experimental setup includes two telescope arms, each with silicon detectors at both ends, to measure the incident and scattered muons. With a length of approximately 5 meters, the telescoping arms provide sufficient leverage for angle reconstruction. A helium or vacuum tube will occupy most of the space between the silicon detectors to control effects from materials and multiple scatterings. The central region of the setup is a Time Projection Chamber (TPC) filled with pressurized hydrogen gas up to 20 bar, serving as an active target and measuring the recoiled proton with kinetic energy varying from 0.5 MeV to 20 MeV. The projected statistical uncertainty with about 260 days of measurement will be better than 0.1\% for the proton electric form factor $G_{E}^{p}$ and better than 0.01 fm for the extracted proton charge radius~\cite{Quintans:2022utc}.


\subsection{The PRES Experiment at Mainz}
The Mainz PRES experiment~\cite{Belostotski:2019qum, Vorobyev:2019hpy} will also utilize an ``active'' hydrogen target, similar to that of the AMBER experiment. The PRES experiment will be conducted in the A2 experimental hall of the Mainz Microtron, using a high-precision 720 MeV electron beam. A TPC filled with high-purity hydrogen gas serves as the target and is capable of detecting recoil protons~\cite{Vorobyev:2019sas}, which features a number of advantages. Firstly, it allows the reconstruction of $Q^{2}$ in $ep$ elastic scattering using proton energy alone, and thus the reconstructed $Q^{2}$ is not sensitive to any pre-vertex energy loss of the electron beam or the uncertainty in the electron beam energy. Additionally, reconstructing $Q^{2}$ using protons significantly suppresses certain radiative effects, such as real-photon bremsstrahlung from the electron line and electron vertex correction, which are the dominant terms for traditional $ep$ scattering experiments that only detect scattered electrons. The scattered electrons will also be detected in coincidence by a Multi-Wire Proportional Chamber (MWPC) based forward tracker. With 45 days of data collection, the experiment aims to collect about 70 million elastic $ep$ events in the $Q^{2}$ range between 0.001 GeV$^{2}$ and 0.04 GeV$^{2}$. The projected relative and absolute uncertainties for differential cross sections are 0.1\% and 0.2\%, respectively. The total uncertainty on $r_{E}^{p}$ is expected to be better than 0.01 fm~\cite{Belostotski:2019qum}.

\subsection{Mainz MAGIX Experiment}

The MAinz Gas Injection target eXperiment (MAGIX) is another upcoming experiment at Mainz~\cite{A1:2021njh}. This experiment will use the Mainz Superconducting Energy Recovery Linac (MESA)~\cite{Hug:2020miu}, which is currently under construction. For MAGIX, the accelerator will be operating in its Energy Recovery Linac (ERL) mode, generating a 1~mA electron beam with energy up to 105~MeV. The experiment will employ an internal cryogenic supersonic gas jet target \cite{A1:2021njh}, which has already been successfully developed and used in the Mainz jet target experiment (discussed in Section~\ref{ch:mainz_jet_exp}). This technique offers various benefits, including reducing pre-vertex radiative effects and multiple scatterings, eliminating backgrounds from target cell windows, and acting as an effectively point-like target. The expected integrated luminosity will be at the order of $10^{35}\rm{cm}^{-2} \rm{s}^{-1}$\cite{A1:2021njh}.

\begin{figure}[H]
\centering
\includegraphics[width=12 cm]{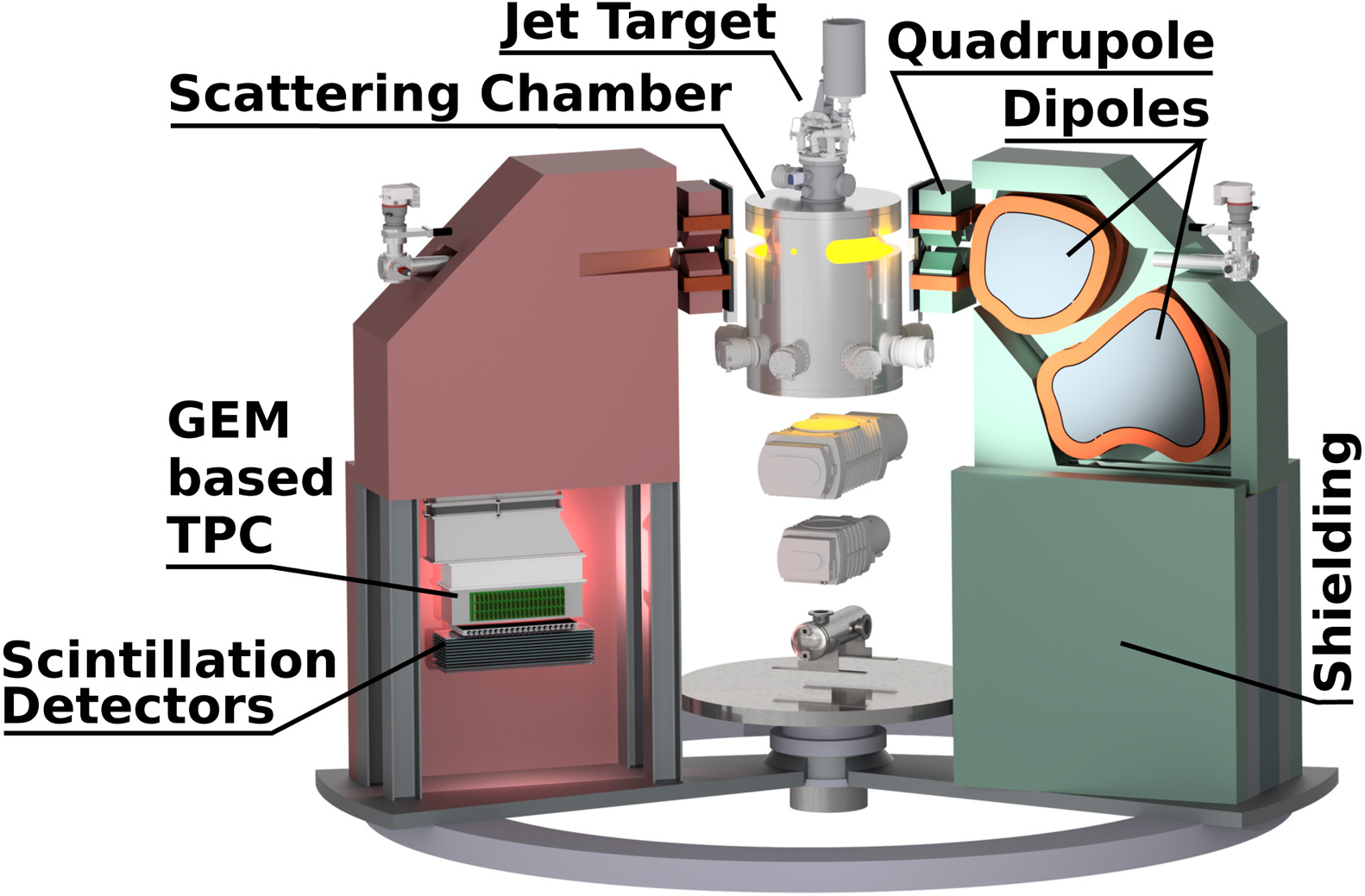}
\caption{The multi-purpose spectrometer system planned for the MAGIX experiment (figure credit: B.~S.~Schlimme \textit{et al.}~\cite{A1:2021njh}).}
\label{fig:magix_spectrometer}
\end{figure}

The MAGIX experiment will utilize a versatile spectrometer system, illustrated in Fig.~\ref{fig:magix_spectrometer}\footnote{This figure is contained in the article that was published by B.~S.~Schlimme \textit{et al.} [A1 and MAGIX], ``Operation and characterization of a windowless gas jet target in high-intensity electron beams,'' Nucl. Instrum. Meth. A \textbf{1013}, 165668 (2021), Copyright Elsevier 2023.}, which consists of a target chamber at the center and a pair of identical magnetic spectrometers linked by a movable vacuum seal to minimize material effects~\cite{Caiazza:2020sda}. Each spectrometer arm has a broad angular coverage (15$^{\circ}$ to 165$^{\circ}$) with respect to the beamline and includes a quadruple magnet followed by two dipole magnets, which bend and focus the final-state particles into the focal plane detectors. The detector array comprises a GEM-based TPC for tracking and high-rate capability, and a trigger veto system consisting of a plastic scintillation detector and a flexible system of additional scintillation detectors and lead absorbers. To further mitigate the material effects, the focal plane detectors adopt an open field-cage design for the TPC~\cite{Caiazza:2020sda}, with an open face pointing towards the spectrometer vacuum chamber. This spectrometer system can achieve a relative momentum resolution of $\sim10^{-4}$ and an angular resolution of approximately 1~mrad.

Low beam energies from the MESA accelerator enable the MAGIX experiment to attain the lowest $Q^{2}$ of approximately $10^{-4}$ GeV$^{2}$, and the highest $Q^{2}$ of about 0.03~GeV$^{2}$. The precision of the proton electric form factor will be mostly below 0.1\%. Additionally, the experiment has a strong sensitivity to the magnetic form factor, with a goal of achieving an order of magnitude improvement in precision in the low $Q^{2}$ region~\cite{Bernauer:2020ont}.

\unskip 
\subsection{ULQ2 Experiment}
The Ultra-Low Q2 (ULQ2) collaboration~\cite{ULQ2, Suda:2022hsm} is planning an electron scattering experiment at the Research Center for ELectron PHoton Science (ELPH) of Tohoku University, Japan. The experiment will use a low-energy electron linac to generate an electron beam with an energy range of 20 to 60 MeV to measure the $ep$ elastic scattering cross section, covering $Q^{2}$ from $3\times10^{-4}$ to $8\times10^{-3}$ GeV$^{2}$. The scattered electrons will be detected by magnetic spectrometers equipped with Single-Sided Silicon Detectors (SSSD) as focal plane detectors. The spectrometers provide a relative momentum resolution of $\sim10^{-3}$ and cover a scattering angle range of $30^{\circ}$ to $150^{\circ}$. To normalize the $ep$ elastic scattering cross section, a CH$_{2}$ target will be employed, and the well-known elastic $e^{12}\rm{C}$ cross section will be measured simultaneously. This experiment also aims to have a high sensitivity on the proton magnetic form factor, and the Rosenbluth separation method will be used to extract $G_{E}^{p}$ and $G_{M}^{p}$ from the cross-section measurements. The expected uncertainty on the extracted $G_{E}^{p}$ is on the order of 0.1\%.

\section{Conclusion and Outlook}
In this paper, we provide an overview of recent developments in the proton charge radius puzzle and low-$Q^{2}$ proton electric form factor measurements from $ep$ elastic scattering experiments. The recent high precision $r_{E}^{p}$ measurement from the PRad experiment~\cite{Xiong:2019umf} is consistent with the $\mu$H spectroscopic results~\cite{Pohl:2010zza, Antognini:2013txn}, but is only about 3$\sigma$ away from the CODATA-2010 recommended value~\cite{Mohr:2012tt}. Furthermore, the lepton scattering community is still trying to understand the difference in the form factor $G_E^{p}$ between the PRad data and previous scattering data within $0.01 < Q^2 < 0.06$ GeV$^2$.

We have also reviewed some recent re-analyses that provide unique insights and seem to be able to bridge the gap by using different functional forms and reduced $Q^{2}$ range. However, there is still no consensus on the best approach to extract $r_{E}^{p}$. Although pseudo-data methods can help to handle systematic uncertainties associated with the fitting procedure, the choice of functional form is strongly kinematic dependent. An agreement within the community may still await future experimental data with higher precision, lower $Q^{2}$ range, and better control of normalization. Fortunately, many new lepton-proton elastic scattering experiments motivated by the radius puzzle are underway, each with unique features and systematics that may shed light on the puzzle and the observed data tension of the proton electric form factor.

Meanwhile, there have been exciting developments on the theoretical side, including modern extraction methods for proton charge radius and \textit{ab-initio} calculations from Lattice QCD. With these dedicated efforts, our understanding of the proton charge radius will be deepened in the next decade. Hopefully, we may have answers for all the remaining puzzles by that time.



\vspace{6pt} 



\authorcontributions{Both authors contribute equally to the draft preparation, review and editing. Both authors have read and agreed to the published version of the manuscript.}

\funding{The work of Chao Peng is supported in part by the U.S. Department of Energy, Office of Science, Office of Nuclear Physics, under Contract No. DE-AC02-06CH11357.}

\institutionalreview{Not applicable.}

\dataavailability{Not applicable.} 

\acknowledgments{The authors would like to thank Haiyan Gao, Ashot Gasparian, Dipangkar Dutta, Jingyi Zhou, Jan Bernauer, and Bj\"{o}rn S\"{o}ren Schlimme for helpful discussion, Steffen Strauch for making the figure for the MUSE experiment, as well as Yi Chen for fruitful discussion concerning the quantum phase-space formalism and the proton charge and magnetization distributions in different reference frames.}

\conflictsofinterest{The authors declare no conflict of interest.} 

\begin{adjustwidth}{-\extralength}{0cm}

\reftitle{References}

\end{adjustwidth}
\end{document}